\documentclass[doublespacing]{elsart}

\usepackage{icarus}

\usepackage{natbib}

\usepackage{longtable}
\usepackage{lscape}
\usepackage{setspace}
\usepackage{lineno}
  
\bibpunct{(}{)}{;}{a}{,}{,}

\usepackage{graphicx}

\usepackage{amssymb}
\usepackage{wasysym}

\newcommand{\BibTeX}{ \textrm{B\kern-.05em\textsc{i\kern-.025em b}\kern-.08em
    T\kern-.1667em\lower.7ex\hbox{E}\kern-.125emX} }

\begin{document}

\begin{frontmatter}



\title{Observational Investigation of the 2013 Near-Earth Encounter by Asteroid (367943) Duende}


\author[lowell]{Nicholas A. Moskovitz},
\author[cu]{Conor James Benson},
\author[cu]{Daniel Scheeres},
\author[umass]{Thomas Endicott},
\author[weiz]{David Polishook},
\author[mit]{Richard Binzel},
\author[mit]{Francesca DeMeo},
\author[nm]{William Ryan},
\author[nm]{Eileen Ryan},
\author[uh]{Mark Willman},
\author[ua]{Carl Hergenrother},
\author[perth]{Arie Verveer},
\author[lco]{Tim Lister},
\author[gso]{Peter Birtwhistle},
\author[salt,mit,psi]{Amanda Sickafoose},
\author[nagoya]{Takahiro Nagayama},
\author[cant]{Alan Gilmore},
\author[cant]{Pam Kilmartin},
\author[psi]{Susan Bennechi},
\author[dtm]{Scott Sheppard},
\author[seti]{Franck Marchis},
\author[not]{Thomas Augusteijn},
\author[not]{Olesja Smirnova}

\address[lowell]{Lowell Observatory, 1400 West Mars Hill Road, Flagstaff, AZ 86001 (U.S.A)}
\address[cu]{University of Colorado Boulder, Boulder, CO 80305}
\address[umass]{University of Massachusetts Boston, Boston, MA 02125}
\address[weiz]{Weizmann Institute of Science, 234 Herzl St. Rehovot 7610001, Israel}
\address[mit]{Massachusetts Institute of Technology, EAPS, Cambridge, MA 02139}
\address[nm]{New Mexico Institute of Mining and Technology, Socorro, NM 87801}
\address[uh]{Institute for Astronomy, University of Hawaii, Hilo, HI 96720}
\address[ua]{University of Arizona, LPL, Tucson, AZ 85721}
\address[perth]{Perth Observatory, Bickley, WA 6076 (Australia)}
\address[lco]{Las Cumbres Observatory Global Telescope Network, Goleta, CA 93117}
\address[gso]{Great Shefford Observatory, Berkshire (England)}
\address[salt]{South African Astronomical Observatory (South Africa)}
\address[nagoya]{Kagoshima University (Japan)}
\address[cant]{University of Canterbury, Mt John University Observatory, Lake Tekapo (New Zealand)}
\address[psi]{Planetary Science Institute, 1700 E. Fort Lowell Rd., Tucson, AZ 85719}
\address[dtm]{Department of Terrestrial Magnetism, Carnegie Institution for Science, Washington, DC 20015}
\address[seti]{SETI Institute, Carl Sagan Center, 189 Bernardo Ave., Mountain View CA 94043}
\address[not]{Nordic Optical Telescope, La Palma (Spain)}

\begin{center}
\scriptsize
Copyright \copyright\ 2019 Nicholas A. Moskovitz
\end{center}


%
%
%
%
%


\end{frontmatter}



\begin{flushleft}
\vspace{1cm}
Number of pages: \pageref{lastpage} \\
Number of tables: \ref{lasttable}\\
Number of figures: \ref{lastfig}\\
\end{flushleft}


\begin{pagetwo}{The 2013 Near-Earth Encounter of Asteroid Duende}

Nicholas A. Moskovitz\\
Lowell Observatory
1400 West Mars Hill Road\\
Flagstaff, AZ 86001, USA. \\
\\
Email: nmosko@lowell.edu\\
Phone: (928) 779-5468 \\

\end{pagetwo}


\begin{abstract}

On 15 February 2013, the asteroid 367943 Duende (2012 DA14) experienced a near-Earth encounter at an altitude of 27,700 km or 4.2 Earth radii. We present here the results of an extensive, multi-observatory campaign designed to probe for spectral and/or rotational changes to Duende due to gravitational interactions with the Earth during the flyby. Our spectral data reveal no changes within the systematic uncertainties of the data. Post-flyby lightcurve photometry places strong constraints on the rotation state of Duende, showing that it is in non-principal axis rotation with fundamental periods of $P_1 = 8.71 \pm 0.03$ and $P_2 = 23.7 \pm 0.2$ hours. Multiple lightcurve analysis techniques, coupled with theoretical considerations and delay-doppler radar imaging, allows us to assign these periods to specific rotational axes of the body. In particular we suggest that Duende is now in a non-principal, short axis mode rotation state with a precessional period equal to $P_1$ and oscillation about the symmetry axis at a rate equal to $P_2$. Temporal and signal-to-noise limitations inherent to the pre-flyby photometric dataset make it difficult to definitively diagnose whether these periods represent a change imparted due to gravitational torques during the flyby. However, based on multiple analysis techniques and a  number of plausibility arguments, we suggest that Duende experienced a rotational change during the planetary encounter with an increase in its precessional rotation period. Our preferred interpretation of the available data is that the precession rate increased from 8.4 hours prior to the flyby to 8.7 hours afterwards. A companion paper by \citet{benson19}  provides a more detailed dynamical analysis of this event and compares the data to synthetic lightcurves computed from a simple shape model of Duende. The interpretation and results presented in these two works are consistent with one another. The ultimate outcome of this campaign suggests that the analytic tools we employed are sufficient to extract detailed information about solid-body rotation states given data of high enough quality and temporal sampling. As current and future discovery surveys find more near-Earth asteroids, the opportunities to monitor for physical changes during planetary encounters will increase.

\end{abstract}



%
%
\section{Introduction \label{sec.intro}}

It has long been understood that the physical properties of small Solar System bodies can be altered when passing within a few radii of a massive planet \citep{roche49}. Numerous theoretical and observational studies suggest that gravitationally-induced physical modification of near-Earth objects (NEOs) can occur during close encounters with the terrestrial planets. The presence of doublet craters and crater chains on the Earth and Venus can be explained by tidal disruption of rubble pile progenitors during close encounters with the Earth \citep{bottke96,richardson98}. Simulations of rubble pile asteroids suggest that close encounters approaching the Earth's Roche limit ($\sim3-5$ Earth radii depending on body density, 1 Earth radius = 6371 km) can result in the redistribution or stripping of surface material, the initiation of landslides, and/or overall changes in body morphology \citep{richardson98,yu14}. The redistribution of surface material can observationally manifest as a change in spectral properties as fresh, unweathered, sub-surface material is excavated \citep{nesvorny05,binzel10}. Such close encounters can also increase or decrease spin rate depending on the details of the encounter, e.g. body morphology, encounter distance, body orientation during encounter, or body trajectory \citep{scheeres00,scheeres04,scheeres05}. The magnitude of these effects scales in various (and sometimes unknown) ways as a function of encounter distance. For the specific case of the asteroid 4179 Toutatis a small ($<1\%$) change in its rotational angular momentum was detected following an Earth-encounter at a distance of four times the Earth moon separation or a distance of over 200 Earth radii \citep{takahashi13}. Despite this foundation of theoretical work that describes the role of planetary encounters in altering the physical properties of asteroids, these effects have never been observed in real-time for encounters between an asteroid and a terrestrial planet. 

We will present here a campaign focused on the near-Earth asteroid 367943 Duende (2012 DA14) to probe for physical changes during its near-Earth flyby in early 2013. As we will demonstrate, Duende is now in a non-principal axis rotation state. The details of how an asteroid enters into non-principal axis rotation are not well understood \citep{pravec05}. For near-Earth objects, tidal torques experienced during planetary encounters can induce such excited rotation states \citep{scheeres00,scheeres04}. The exact encounter distance at which non-principal axis rotation can be induced depends sensitively on the orientation of the asteroid relative to the planet at the time of the flyby and the initial rotation rate of the asteroid. Percent level changes in spin rate can be induced for encounters at distances of many tens of Earth radii \citep{scheeres00}. Once in a non-principal axis state, energy is dissipated through internal stress causing a damping of the excited rotation and evolution into a state of constant rotation about the axis of maximum moment of inertia, i.e. principal axis rotation. The characteristic time scale for damping of non-principal axis rotation is dependent on internal properties like rigidity and efficiency of energy dissipation, rotation rate, and body size \citep{burns73,harris94}. For the population of known tumblers \citep{warner09} smaller than 100 meters (relevant to NEOs and Duende), cohesionless damping timescales range from greater than the age of the Solar System to under 1 Myr \citep{sanchez14}. Open questions remain regarding how internal structure, e.g. cohesionless rubble piles vs. coherent bodies, affects these damping timescales \citep{sanchez14}. 

Asteroid discovery surveys are currently yielding more than 100 new NEOs every month and are increasingly finding smaller bodies as completeness is reached for large objects \citep{galache15}. The shift towards increasingly smaller NEOs means that these objects must be at smaller geocentric distances to reach survey detection limits. Current surveys are finding several objects every month that pass within at least 1 lunar distance of the Earth (1 LD = 60 Earth radii = 0.0026 AU). Typically objects that undergo these near-Earth encounters are discovered only days before closest approach, requiring rapid response capabilities to conduct telescopic observations. However, this was not the case with the discovery of Duende.

On UT 23 February 2012 the La Sagra Sky Survey discovered the approximately 40 meter near-Earth asteroid 367943 Duende. Within days of discovery it was realized that this asteroid would experience a close encounter with the Earth a year later on 15 February 2013 around 19:25 UT (JD 2456339.3090). This flyby would occur at an altitude of roughly 27,700 km or 4.2 Earth radii, inside the orbital ring of geosynchronous satellites and just outside the Earth's Roche limit. This would be the 8th closest encounter ever recorded between an asteroid and the Earth, and the first time such an event was known about more than a few days in advance. The recovery of the asteroid a year after discovery in January 2013 \citep{moskovitz13} and the subsequent refinement of the object's orbit enabled characterization observations during this unusual event.

To investigate the possibility of gravitationally induced changes to Duende's physical properties we conducted an extensive campaign of spectroscopic and photometric observations immediately after discovery in 2012 and one year later surrounding the close encounter. The primary focus of our observational campaign involved monitoring the rotational state of the asteroid by measuring photometric lightcurves. We present the result of the observations in \S\ref{sec.obs}. The lightcurve photometry clearly indicates that Duende is in a non-principal axis rotation state. In \S\ref{sec.analysis} we perform detailed analyses of the lightcurve data, namely least-squares fitting and computation of Fourier power spectra, on the data collected prior to and following the near-Earth flyby. This enables physical interpretation of the spin state of the body and a comparison of the pre and post-flyby rotation rates (\S\ref{sec.comparison}). A companion paper \citep{benson19} theoretically explores the solid body dynamics of Duende during the planetary encounter and finds similar results to what are presented here. These dynamical models coupled with feasibility arguments founded on our data analysis suggests that Duende may have experienced rotational changes during the planetary encounter in 2013.  The implications of this work are discussed in Section \ref{sec.disc}.

%
%
\section{Observations and Data Reduction \label{sec.obs}}

Our observational campaign involved photometry and spectroscopy at visible wavelengths. We also leverage Doppler delay radar imaging and radar speckle tracking results \citep{benner13}. The visible spectra were obtained to search for spectral changes attributable to seismically induced resurfacing that would reveal fresh, unweathered, subsurface material \citep{binzel10}. The lightcurve photometry was obtained to search for rotationally induced changes or mass loss during the planetary encounter. The radar observations constrain the rotation state and morphology of the asteroid.

%
\subsection{Visible Spectroscopy \label{sec.spec}}

\begin{figure}[]
\begin{center}
\includegraphics[width=14cm]{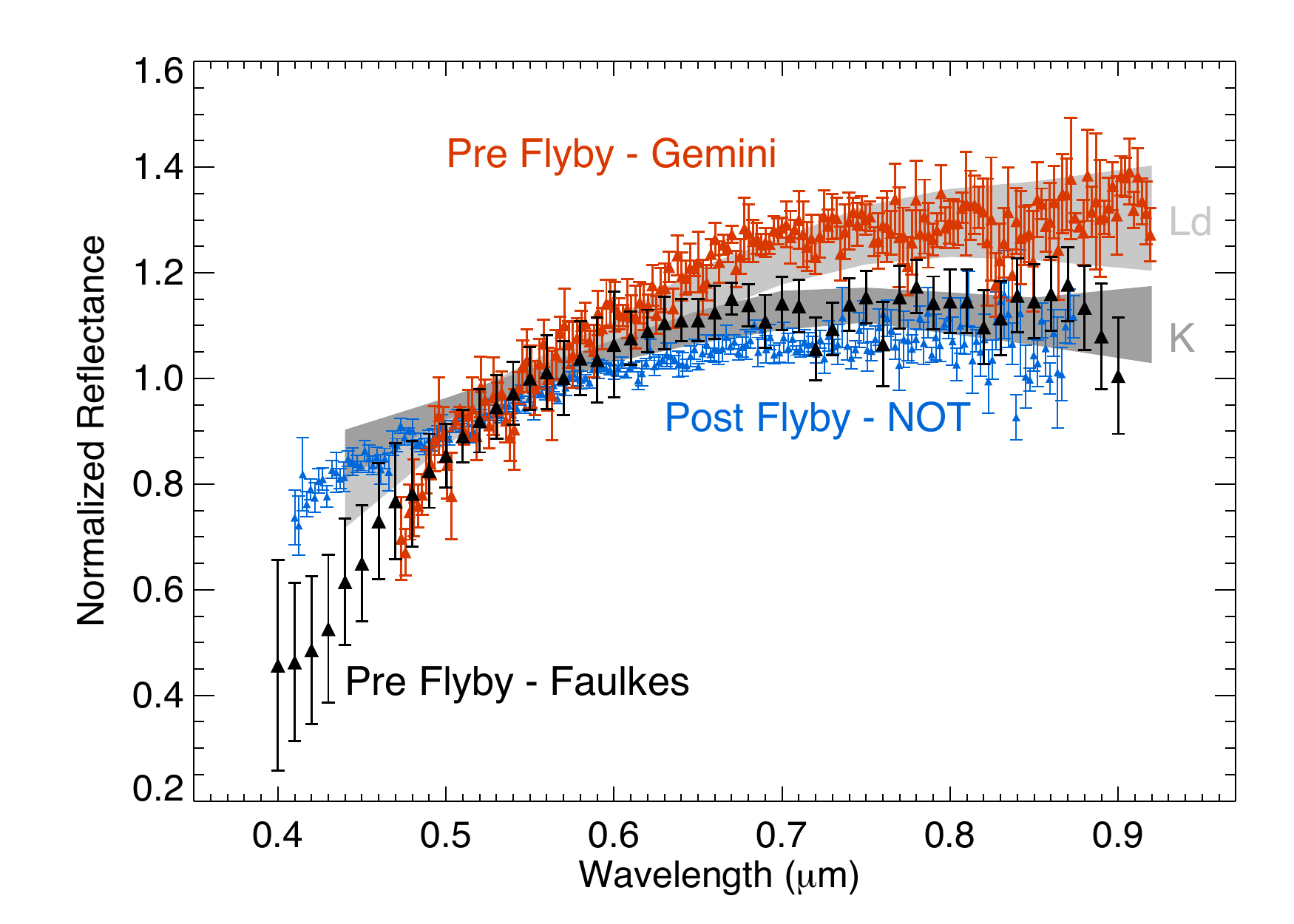}
\end{center}
\caption[]{Normalized visible spectra of Duende including pre-flyby data from the discovery epoch in 2012 (Gemini-GMOS, red), pre-flyby data about 9 hours before closest approach (Faulkes-FLOYDS, black), and post-encounter data taken two days after the flyby (NOT-ALFOSC, blue). Taxonomic envelopes from the system of \citet{bus02} are shown in grey in the background for Ld- and K-types, an L-type classification is intermediate to these two. The variability between spectra is most likely systematic in nature and not due to spectral changes of the asteroid following the planetary encounter.
} 
\label{fig.spec}
\end{figure}

Three visible wavelength ($\sim0.4-0.9~\mu m$) spectra are presented here, two taken prior to the flyby and one afterwards (Table \ref{tab.spec}, Figure \ref{fig.spec}). The spectral data were obtained with GMOS (Gemini Multi-Object Spectrograph) at the Gemini North 8m on Mauna Kea in Hawaii, FLOYDS (Folded Low-Order Yte-pupil Double-dispersed Spectrograph) on the Faulkes South telescope at Siding Springs Observatory in Australia, and with ALFOSC (Alhambra Faint Object Spectrograph and Camera) at the Nordic Optical Telescope (NOT) on La Palma in the Canary Islands. In all cases these spectra and their errors represent an error-weighted average of multiple individual exposures.

\begin{table}[]
\tiny
\begin{tabular}{lllccccl}
\hline 
\hline
Telescope 				& Location	 			& Instrument	 	& UT Date		 	& V-mag		& Phase Angle		& Exposures			& Solar Analog \\
\hline
\multicolumn{8}{l}{\bf{Discovery Epoch, March 2012}}\\
Gemini North 8m	 		& Mauna Kea, HI 			& GMOS		 	& 2012.03.02	 	& 19.9		& 60$^\circ$		& 12 x 500s			& SA107-998 \\
\hline
\multicolumn{8}{l}{\bf{Pre-Flyby, February 2013}}\\
Faulkes South 2m	 		& Siding Spring, Australia	 	& FLYODS	 	& 2013.02.15	 	& 14.1		& 109$^\circ$		& 2 x 180s			& SA98-978 \\
\hline
\multicolumn{8}{l}{\bf{Post-Flyby, February 2013}}\\
Nordic Optical Telescope 2.5m	& La Palma, Canary Islands 	& ALFOSC	 	& 2013.02.17	 	& 15.8		& 84$^\circ$		& 7 x 300s			& HD245 \\
\hline
\hline
\end{tabular}
\caption{Observational Summary of Duende Spectroscopy}
\label{tab.spec}
\end{table}%

The GMOS instrument at Gemini-North was employed 10 days after the discovery of the asteroid in 2012 following a successful request for director's discretionary time. The instrument was configured with the R150 grating, a 2.0" slit and a GG455 blocking filter, producing a useful spectral range from 0.48 to 0.92 microns at a resolution per channel of approximately 0.35 nm. Three sets of 4 x 500s exposures were obtained with a different grating angle for each set. The three grating angles yielded central wavelengths of 0.69, 0.7 and 0.71 microns and minimized the effects of gaps between the three detectors that populate the focal plane. The GMOS slit was aligned to the default instrument position angle of 90$^\circ$. As the observed airmass of the targets was between 1.24 and 1.32, differential refraction was negligible, particularly with a 2" slit, and thus did not necessitate aligning the slit with the parallactic angle. Reduction of these data employed the Gemini IRAF package. These observations were made challenging by the faintness of the object (V$\sim$20) coupled with its moderate non-sidereal rates of 3"/min. The individual reduced spectra showed significant variability in slope at the $\sim$15\% level. The combination of all 12 GMOS spectra in Figure \ref{fig.spec} shows error bars that solely account for the average measured signal-to-noise per pixel and do not represent this slope variability. The cause of this significant slope variability is not clear. Weather conditions were generally favorable: Gemini metrics of image quality and cloud cover were in the 70-80th percentile and 50-70th percentile respectively. The object appeared well aligned on the slit in the acquisition frames, though no additional images of the slit were taken to confirm alignment once the 12 x 500s spectral sequence began.  The measured variability was largely tied to the last 4 exposures with the grating in the 0.69 micron setting. These exposures were taken at the lowest airmass. There were no clear indications in these spectral images of background field star contamination, however inspection of finder charts do indicate possible contamination from at least one faint field star during these last few exposures. It is possible this could be the cause of the slope variability, but that is difficult to confirm.

A second pre-flyby spectrum was obtained using the FLOYDS instrument on the Faulkes-South telescope on the day of the flyby. FLOYDS has a fixed configuration that produces spectra from approximately 0.4 to 0.9 microns. Spectra are captured in two orders, a red order covering 0.6 to 0.9 microns at a per-pixel resolution of 0.35 nm and a blue order covering 0.4 to 0.6 microns at a per-pixel resolution of 0.17 nm. A 2" slit was used for these observations. We re-binned and spliced the red and blue orders to increase signal-to-noise, producing a final spectrum at a resolution of 0.01 microns (Figure \ref{fig.spec}). The high non-sidereal motion of the asteroid during these observations ($>$80 "/min) made acquisition sufficiently challenging that only two spectra were successfully captured. In one case the object clearly drifted off of the slit during the exposure as indicated by significantly reduced counts. As such the slope of the final spectrum may have been affected. The slit was aligned with the parallactic angle for these observations.

The ALFOSC instrument at the NOT was used 2 days after the flyby on 17 February 2013. The instrument was configured with the \#11 grism and a 1.8" slit, producing a useful spectral range from 0.40 to 0.87 $\mu m$ at a resolution per channel of approximately 0.47 nm. Reduction of these data employed the IRAF {\it specred} package \citep{tody86}. The observing conditions on this night were not good with high humidity, clouds, and poor seeing. Individual spectra showed variability in slope at the $\sim$10\% level. The combination of all 7 ALFOSC spectra in Figure \ref{fig.spec} shows error bars that solely account for the average measured signal-to-noise per pixel and do not represent this slope variability. The ALFOSC slit was aligned with the parallactic angle.

The spectra are best fit with an Ld-, K-, or L-type classification in the \citet{bus02} system. These taxonomic assignments are consistent with the findings of \citet{deleon13} and \citet{takahashi14}. The composition of these taxonomic types is not well understood, though suggested connections to carbonaceous chondrite meteorites have been made \citep{bell88}. This ambiguous composition complicates our search for spectral changes that might be attributed to exposure of fresh, sub-surface material. Carbonaceous chondrites are not generally associated with traditional lunar-style space weathering that causes an increase in spectral slope and a decrease in absorption band depth  \citep{hapke01}. In fact, some studies find that carbonaceous meteorites get spectrally bluer during simulated space weathering processes \citep{lantz13}. Thus, the composition of Duende may spectroscopically weather in a manner that is not currently well understood. Regardless of this ambiguity, we attribute the observed spectral slope variation (Figure \ref{fig.spec}) to observational systematics -- e.g. improper alignment of the asteroid on the slit, airmass differences between the asteroid and calibration solar analog star -- and not physical changes on the asteroid's surface. Though there are differences in the slopes of the final spectra presented here (Figure \ref{fig.spec}), the variation amongst individual exposures from a single instrument generally span an appreciable fraction of this slope range. Furthermore, an instrumental or observational cause of slope variation is supported by the pronounced difference in spectral slope between the two pre-flyby spectra, taken before any potential surface perturbation could have occured. Even though these observations were taken across a range of phase angles (Table \ref{tab.spec}), there is no clear signature of slope variability that could be attributed to phase reddening \citep{sanchez12}. Therefore we conclude that this slope variability is systematic in nature and not due to spectral changes intrinsic to the asteroid. This suggests that the surface was not significantly perturbed during the flyby and thus did not reveal unweathered subsurface material, or the composition of Duende is insensitive to space weathering effects at visible wavelengths, or any spectral changes were below the sensitivity of our measurements. This result is fully consistent with the non-detection of near-infrared color changes during the flyby \citep{takahashi14}.

%
\subsection{Photometry \label{sec.phot}}

\begin{figure}[]
\begin{center}
\includegraphics[width=14cm]{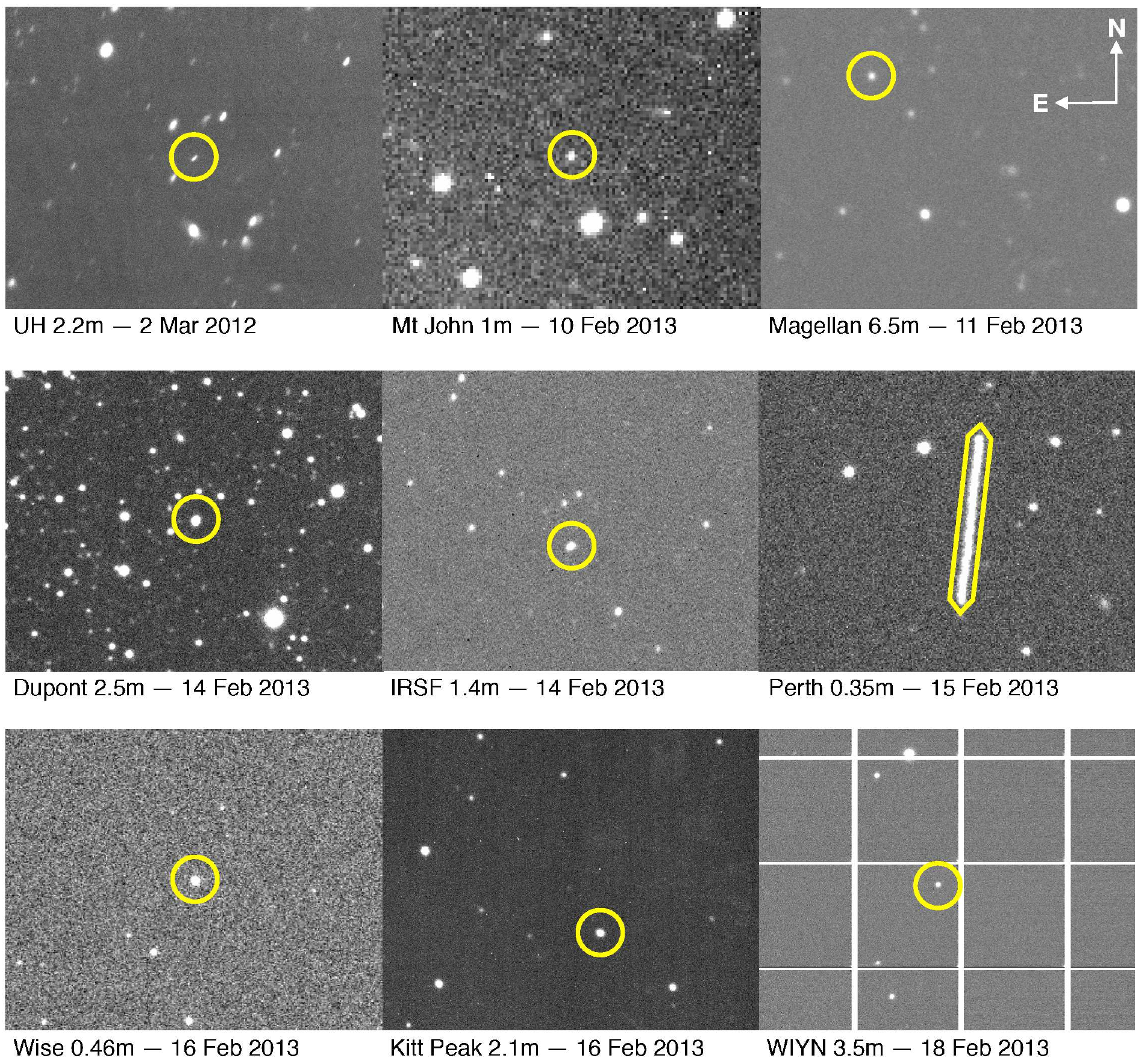}
\end{center}
\caption[]{Example images from several of the instruments used in the photometry campaign. Each field is approximately 3' wide and all are orientated with North up, East left as indicated by the compass in the upper right. Duende is circled in each frame, except for the image from the Perth 0.35m, where the polygonal photometric aperture is shown.
} 
\label{fig.images}
\end{figure}

\begin{figure}[]
\begin{center}
\includegraphics[width=14cm]{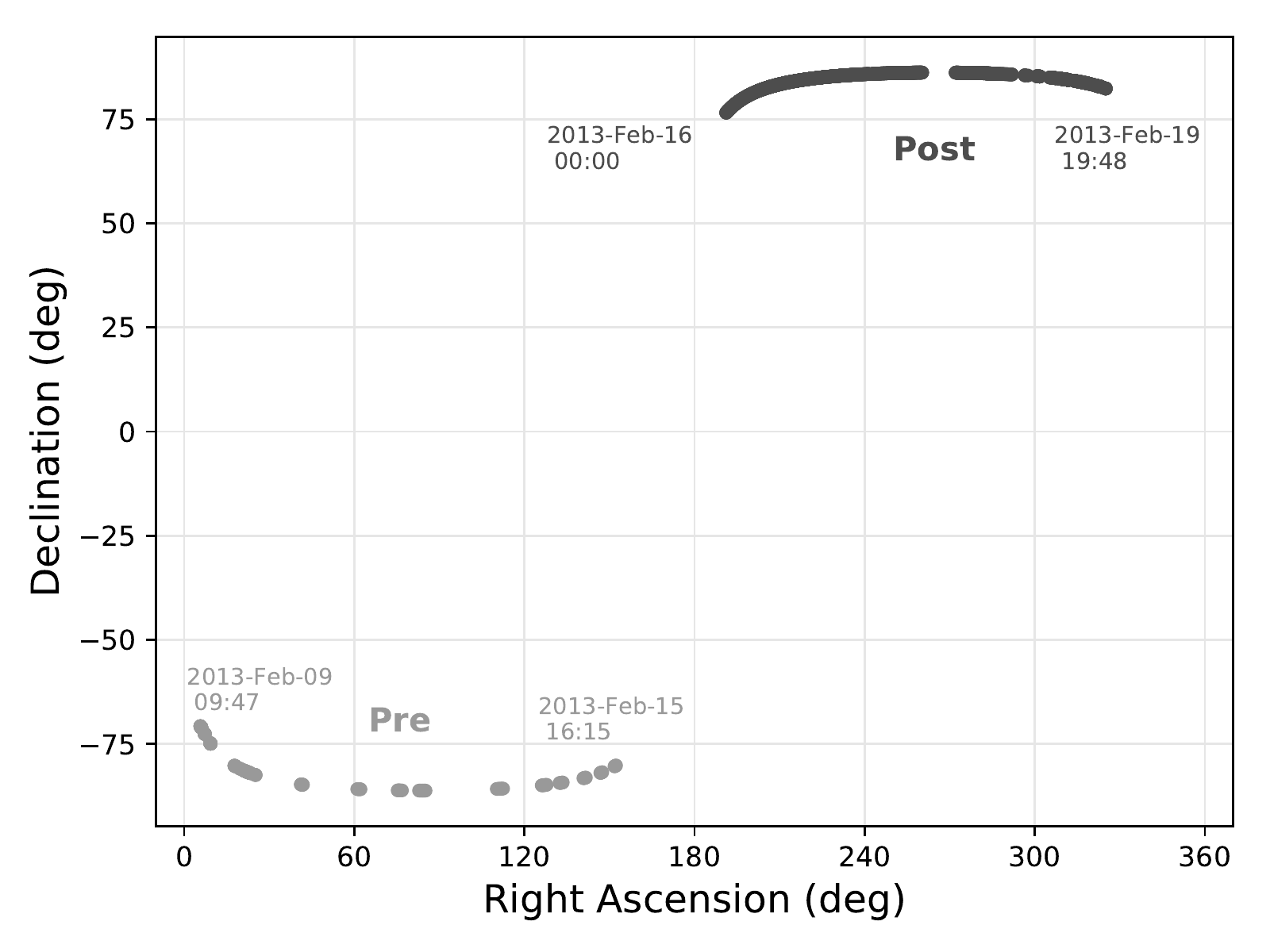}
\end{center}
\caption[]{Geocentric coordinates surrounding the dates of closest approach. Each point represents one of our lightcurve observations, both pre (light grey) and post (dark grey) encounter are included. The start and end times of the pre and post-flyby sequences are shown.
} 
\label{fig.obs_geo}
\end{figure}

The primary focus of our observational campaign involved measuring rotational lightcurves via broadband photometry. These observations were conducted in March 2012 following the discovery of Duende and in February 2013 on either side of the close encounter. In total these observations incorporated data from 15 different observatories. Examples of images from a number of these observatories are shown in Figure \ref{fig.images}. The orbital geometry of the flyby was such that the asteroid approached from a large negative declination of approximately -75$^\circ$ and receded at a large positive declination around +80$^\circ$. Thus, pre-flyby characterization required observations conducted from the southern hemisphere, while post-flyby observations were conducted from the North (Figure \ref{fig.obs_geo}). These observations spanned effectively the full extent of the observable sky: more than 300$^\circ$ in right ascension and more than 150$^\circ$ in declination. In the hours surrounding the flyby, a rapidly changing viewing geometry and the lack of a detailed shape model would have made it difficult to link the body's physical rotation state to measured lightcurves. Thus, our strategy was to compare rotation state before and after the flyby to look for evidence of spin changes. Details of the lightcurve observations are summarized in Table \ref{tab.phot} and each observatory/instrument combination is described below. In general reduction of our photometric data followed standard aperture photometry techniques in IRAF following the methods of \citet{moskovitz12}, though several exceptions are noted. All data were transformed to V-band magnitudes \citep{dandy03} based on reference magnitudes of on-chip field stars. The temporal overlap of lightcurve segments from multiple observatories (Table \ref{tab.phot}) provides confirmation that our magnitude calibration techniques are reasonable and accurate to $<0.1$ magnitudes. For analysis purposes all data have been corrected for geocentric range and phase angle using an ephemeris generated with the JPL Horizons system so that we only work with differential magnitudes relative to the ephemeris predictions.

\begin{landscape}
\begin{table}[]
\scriptsize
\begin{tabular}{llccccl}
\hline 
\hline
		 				& 			 			& 			 	& CCD 		 	& Plate 			& 				& Observation Time \\
Telescope 				& Location	 			& Instrument	 	&  Size		 	& Scale			& Filter			& (JD-2455000) \\
\hline
\multicolumn{7}{l}{\bf{Discovery Epoch, March 2012}}\\
University of Hawaii 2.2m 		& Mauna Kea, HI 			& OPTIC		 	& 4k x 4k		 	& 0.137"			& I				& 988.90347 - 989.15692 \\
\hline
\multicolumn{7}{l}{\bf{Pre-Flyby, February 2013}}\\
Mt. John Observatory 1m 		& Lake Tekapo, New Zealand 	& Apogee CCD	 	& 1k x 1k		 	& 1.3"			& Open			& 1332.89688 - 1335.00426 \\
Magellan Baade 6.5m 		& Las Campanas, Chile	 	& IMACS		 	& 8k x 8k		 	& 0.2"			& WB4800-7800	& 1334.54623 - 1334.57861 \\
LCOGT 1m		 		& Cerro Tololo, Chile		 	& SBIG CCD		& 4k x 4k		 	& 0.232"			& PanSTARRS-w	& 1334.61609 - 1334.65641 \\
DuPont 2.5m		 		& Las Campanas, Chile	 	& SITe2k CCD		& 2k x 2k		 	& 0.259"			& SDSS r'			& 1337.52066 - 1337.54869 \\
IRSF 1.4m		 		& SAAO, South Africa	 	& SIRIUS			& 1k x 1k		 	& 0.46"			& J				& 1338.40220 - 1338.42668 \\
Faulkes South 2m	 		& Siding Spring, Australia 		& Spectral			& 4k x 4k		 	& 0.3"			& VR				& 1338.90788 - 1338.97700 \\
Perth Observatory 0.35m		& Perth, Australia		 	& RCOP			& 2k x 1.5k	 	& 0.66"			& Open			& 1339.05211 - 1339.17737 \\
\hline
\multicolumn{7}{l}{\bf{Post-Flyby, February 2013}}\\
Wise Observatory 0.46m 		& Tel Aviv, Israel		 	& SBIG CCD	 	& 2k x 1.5k	 	& 1.1"			& Open			& 1339.50033 - 1339.65758 \\
VATT 1.8m		 		& Mt. Graham, Arizona	 	& VATT4k		 	& 4k x 4k		 	& 0.38"			& R				& 1339.59879 - 1339.74846 \\
Kitt Peak 2.1m				& Kitt Peak, Arizona		 	& T2KB		 	& 2k x 2k		 	& 0.3"			& SDSS g'		& 1339.97859 - 1340.05798 \\
Great Shefford Observatory	& West Berkshire, England 	& Apogee CCD	 	& 1k x 1k		 	& 2.2"			& R				& 1340.38243 - 1340.51431 \\
Magdalena Ridge 2.4m		& Socorro, New Mexico	 	& 2k Imager	 	& 2k x 2k		 	& 0.53"			& R				& 1340.63117 - 1341.73574 \\
WIYN 3.5m				& Kitt Peak, Arizona		 	& pODI		 	& 12k x 12k	 	& 0.11"			& SDSS r'			& 1341.78420 - 1342.05401 \\
\hline
\hline
\end{tabular}
\caption{Observational Summary of Duende Photometry Surrounding the Close Encounter at JD 2456339.3090}
\label{tab.phot}
\end{table}%
\end{landscape}

{\it University of Hawaii 2.2m:} We employed the Orthogonal Parallel Transfer Imaging Camera (OPTIC) on the University of Hawaii 2.2 meter telescope to obtain Cousins I-band photometry of Duende within a week of its discovery. OPTIC consists of two 2k x 4k Lincoln Lab detectors covering a field of view roughly 9.3' x 9.3'. The telescope was tracked at half the non-sidereal rate of the asteroid to minimize elongation of the field stars and asteroid in each exposure such that circular aperture photometry could be performed. We used an exposure time of 120 seconds. Reduction of these data employed standard IRAF routines.

{\it Mt John Observatory 1m:} We employed an Apogee F6 1k x 1k KAF1001E CCD with 24 $\mu m$ pixels and a plate scale of 0.64". The detector was binned 2 x 2 and exposure times were set to 60 seconds. The telescope was guided at sidereal rates allowing the asteroid to move through the fixed 11' x 11' field of view. Reduction of these data employed standard IRAF tools.

{\it Magellan Baade 6.5m:} Observations were conducted with IMACS \citep{dressler11} at the Magellan Baade 6.5 m telescope at Las Campanas Observatory in Chile. IMACS was operated using the f/2 camera, which has a 27.50 field-of-view covered by a mosaic of eight 2k x 4k CCDs with plate scales of 0.2"/pixel. Exposure times of 7-10 seconds and a broadband filter with spectral response from 0.48 - 0.78 $\mu m$ were used. Reduction of these data employed standard IRAF tools.

{\it Las Cumbres Observatory Global Telescope Network (LCOGT) 1m:} We employed a 4k x 4k SBIG CCD with a 0.232" plate scale covering a 15.8' field of view and an exposure time of 180 seconds. The frames were binned 2 x 2. Data from this observatory and the Faulkes South 2m (described below) were pre-processed through the LCOGT pipeline \citep{brown13} to perform bias and dark-subtraction, flat fielding, and to determine an astrometric solution against the USNO-B1.0 \citep{monet03} for the 2m and the UCAC-3 \citep{zacharias10} catalog for the 1m.  The 1m photometry were measured using SExtractor \citep{bertin96} to fit elliptical apertures to the trailed, elongated asteroid images.

{\it DuPont 2.5m:} Observations were performed at the DuPont 2.5 m telescope at Las Campanas Observatory with a SITe2k CCD. The SITe2k at DuPont is a 2k x 2k CCD with an 8.85' field-of-view and a plate scale of 0.259"/pixel. Exposures of 60 seconds were obtained. Reduction of these data employed standard IRAF tools.

{\it Infrared Survey Facility (IRSF) 1.4m:} Observations were obtained with the Simultaneous three-color Infrared Imager for Unbiased Surveys (SIRIUS) at the IRSF located in Sutherland, South Africa. SIRIUS consists of three 1k x 1k Hawaii HgCdTe arrays (for simultaneous J, H and K band imaging) with plate scales of 0.453"/pixel, each covering a 7.7' field of view.

 {\it Faulkes South 2m Telescope:} We employed the Spectral instrument which contains a 4k x 4k camera for imaging. The plate scale of Spectral is 0.152" and the field of view is 10.5'. Frames were binned 2 x 2. Exposure times ranged from 180 seconds on UT 9 February 2013 to 7 seconds on 15 February 2013. The frames were processed in Astrometrica using the UCAC-3 catalog for astrometric reference and then positions and magnitudes were determined.

{\it Perth Observatory R-COP 0.35m:} Observations were conducted on the day of the flyby with the 0.35m R-COP telescope at Perth Observatory using a 2k x 1.5k SBIG CCD with 0.66" pixels and a 24.7' x 16.5' field of view. All images were obtained unfiltered. Exposure times ranged from 10 seconds at the onset (when the object was moving slowest) to 1 second about three hours prior to closest approach when the object was moving at non-sidereal rates in excess of 500 "/minute. These were the only data for which we did not employ standard circular or elliptical aperture photometry techniques. The high non-sidereal rate of the asteroid in these images resulted in trailing. As such we employed polygonal aperture photometry with the IRAF {\it polyphot} routine. Field stars in these images were examined to determine the seeing; the width of the polygonal aperture along the short (untrailed) axis of the asteroid was interactively defined to equal approximately twice the local seeing. The polygonal apertures were defined with six points, three at each end of the trailed asteroid. The background annulus was defined with a 60-pixel inner radius and a 30-pixel width. 

{\it Wise Observatory 0.46m:} The 0.46m at Wise Observatory in Israel \citep{brosch08} was employed about 4.5 hours after closest approach. The instrument was a 2k x 1.5k SBIG CCD with 1.1" pixels and a 40.5' x 27.3' field of view. Images were obtained unfiltered. The telescope pointing was offset between sets of exposures to keep up with the fast motion of the asteroid. This was conducted automatically throughout the night of 2013 February 15 by programming the telescope to follow ephemeris coordinates from the Minor Planet Center (MPC). IRAF's {\it phot} function with an aperture of 4 pixels was used for measuring the photometry. The differential magnitude of the asteroid was calibrated based on comparison to hundreds of on-chip field stars. The brightness of these stars remained constant to within $\pm$0.02 mag. See \citet{polishook09} for details of the reduction algorithm. Nine different fields-of-view were used to track the asteroid over the course of about 4.5 hours. This resulted in nine lightcurve segments that were stitched together to form a continuous lightcurve.

{\it Vatican Advanced Technology Telescope (VATT) 1.8m:} The VATT 1.8m was operated with the VATT 4k STA CCD. The CCD covers a 12.5' square field of view at 0.188 "/pixel. Exposure times ranged from 2 to 5 seconds and the telescope was tracked at the non-sidereal rates of the asteroid. Trailing of the field stars was minor thus circular aperture photometry was possible.

{\it Kitt Peak 2.1m:} The T2KB Tektronix 2k x 2k CCD was employed at the Kitt Peak 2.1m. The plate scale of this instrument is 0.3"/pixel and the un-vignetted field of view of 10.2' x 9.4'. We obtained 5 second exposures and guided the telescope at sidereal rates allowing the asteroid to move through fixed star fields. Reductions and photometry followed standard IRAF protocols.

{\it Great Shefford Observatory 0.4m:} The 0.4m Schmidt-Cassegain telescope at Great Shefford Observatory in Berkshire, England was used with a 1k x 1k E2V CCD. The plate scale of the CCD is 1.1 "/pixel and produces a vignetted 18' circular field of view. Exposure times of 15 to 30 seconds were used. Reduction of the photometry was performed using Astrometrica referenced to the PPMXL \citep{roeser10} catalog.

{\it Magdalena Ridge Observatory (MRO) 2.4m:} The MRO 2.4m was employed with a 2k x 2k CCD with a 0.53"/pixel plate scale that images an 18' square field of view. Observations were made tracking on Duende while periodically shifting to a nearby comparison field. Seeing was typically 1.0-1.2" on UT 17 February 2013 and 1.5-1.9" on 18 February 2013. Data were measured using IRAF's apphot package with 12 pixel and 16 pixel apertures on 17 February and 18 February respectively (large apertures allowed for seeing variations). The comparison fields were calibrated using Landolt standards and differential photometry was performed using a custom program assuming a V-R of 0.47 for the target, roughly consistent with Duende's L-type taxonomic classification \citep{dandy03}.

{\it WIYN 3.5m:} The WIYN 3.5m was employed with the partial One Degree Imager \citep[pODI;][]{harbeck14} which is a mosaic of multiple 4k x 4k STA orthogonal transfer CCDs. These chips have a plate scale of 0.11"/pixel and each covers an 8' square field of view. We employed the central 9 CCDs to produce an overall field 24' x 24'. The wide field-of-view of pODI allowed us to use a single pointing for the entire night of UT 17 February 2013. This enabled use of a single set of reference field stars for measuring the differential magnitude of the asteroid. Many images were excluded from analysis because of the proximity of the asteroid to pODI's numerous chip and cell boundaries. Reduction of the data employed a custom set of python routines (courtesy R. Kotulla). Photometry was performed using IRAF's apphot package with an aperture of 15 pixels, and a background annulus with radius 30 pixels and width of 25 pixels.

In addition to the photometry obtained as part of our campaign, we used publicly available data from \citet{gary13} and a few select measurements submitted to the MPC. We only included sets of MPC data for which there were 3 or more observations within a single night and when the data did not show large ($>$0.2 magnitude) random fluctuations (Table \ref{tab.mpc}). These criteria removed the vast majority of the more than 1000 individual MPC observations of Duende, adding only 11 data points to our pre-flyby lightcurve and 161 to our post-flyby curve.

Our post-flyby images showed no evidence for mass loss or mass shedding. No difference was seen in the width of Duende's point-spread function (PSF) compared to that of field stars in the same images, though a detailed search for faint extended features \citep[e.g.][]{sonnett11} was not conducted. The lack of clear signatures for mass loss is consistent with post-flyby radar data that were sensitive enough to resolve, but did not detect, the escape of sub-meter-scale fragments from the surface \citep{benner13}.

\begin{table}[]
\scriptsize
\begin{tabular}{lll}
\hline 
\hline
UT Date 				& Observatory	 					& MPC Code \\
\hline
\multicolumn{3}{l}{\bf{Pre-Flyby}}\\
2013.02.14	 		& Cerro Tololo Observatory			& 807 \\
2013.02.15	 		& Arcadia							& E23 \\
2013.02.15	 		& Murrumbateman					& E07 \\
\hline
\multicolumn{3}{l}{\bf{Post-Flyby}}\\
2013.02.16	 		& Kurihara						& D95 \\
2013.02.16	 		& Pulkova							& 084 \\
2013.02.16	 		& Montevenere Observatory			& C91 \\
2013.02.16, 2013.02.19	& Observatorio La Vara, Valdes			& J38 \\
2013.02.16	 		& Bisei Spaceguard Center - BATTeRS	& 300 \\
2013.02.17	 		& Observatoire de Dax				& 958 \\
2013.02.17	 		& Kourovskaya						& 168 \\
2013.02.18	 		& iTelescope Observatory, Mayhill		& H06 \\
2013.02.18	 		& Sternwarte Mirasteilas, Falera		& B67 \\
\hline
\hline
\end{tabular}
\caption{Summary of MPC Data Used in lightcurve Fits}
\label{tab.mpc}
\end{table}%

%
\subsection{Radar Constraints \label{sec.radar}}

Post-flyby radar imaging and speckle tracking \citep{benner13} provide important constraints, that in conjunction with the lightcurve analysis, allow us to better understand the post-flyby shape and spin state. The delay-Doppler images are consistent with a multi-hour rotation state around 8 hours, but they do not provide a sufficiently long interval to uniquely constrain any non-principal axis rotation. As we show in the following sections, the lightcurve photometry clearly shows that Duende is in non-principal axis rotation. The radar imaging suggests that the body is in a relatively flat spin, which is most consistent with non-principal short-axis mode (SAM) rotation as opposed to a long-axis mode (LAM) precession state \citep{kaasalainen01}, of which asteroid 4179 Toutatis is a well known example \citep{ostro95,hudson95}. Other NEOs with SAM rotation states include 99942 Apophis \citep{pravec14} and (214869) 2007 PA8 \citep{brozovic17}. Radar speckle tracking from the Very Large Array (VLA) constrain the amplitude of non-principal axis wobble (i.e. the long axis nutation angle) to $<30^\circ$, which is also consistent with a SAM state. A more detailed discussion of the body dynamics is presented in \citet{benson19}. The highest resolution images from the Goldstone array in California achieved 3.75m per pixel resolution and determined that the asteroid is highly elongated (approximately 40 $\times$ 20 meters), though no surface features were resolved.

%
%
\section{Lightcurve Analysis \label{sec.analysis}}

As discussed above, lightcurve photometry was obtained during three windows: in 2012 during the discovery apparition, in 2013 before the time of closest approach (19:25 UT 15 February 2013), and in 2013 following the flyby. The radar data were collected in the days following closest approach. We focused our observations and analysis on data taken more than 3 hours before and 3 hours after the time of closest approach. This restriction was an intentional part of our experimental design and had a number of important benefits. First, avoiding the time of closest encounter ensured that the non-sidereal rates of the asteroid were an order of magnitude lower than at their peak. This made the observations and reductions simpler because significant trailing of the asteroid was avoided and a relatively small number of fields could be used for differential photometry against background field stars. Second, this strategy ensured that the viewing geometry did not change significantly during the windows of observation. This can be quantified in terms of the solar phase angle, which changed by $<25^\circ$ during the inbound observing window and $<20^\circ$ outbound. Third, this strategy was based on models suggesting that the majority of gravitationally-induced changes in rotation state occur within a few hours of closest approach \citep{scheeres00}. Thus, our observing windows served to isolate the pre- and post-flyby rotation states of the asteroid, independent of any potential changes. Fourth, conducting our observations in these discrete windows meant that the difference in sidereal versus synodic rotation periods caused insignificant phase shifts of $<$10 seconds. Here we describe the resulting lightcurves and the fitting of those data to derive constraints on the rotational state of the asteroid. Specifically, we perform both a minimum-RMS analysis and a fast Fourier transform (FFT) analysis on each of the data sets to facilitate comparison of the pre and post-flyby rotation states.

%
\subsection{Post-Flyby \label{sec.post}}

\begin{figure}[t]
\begin{center}
\includegraphics[width=14cm]{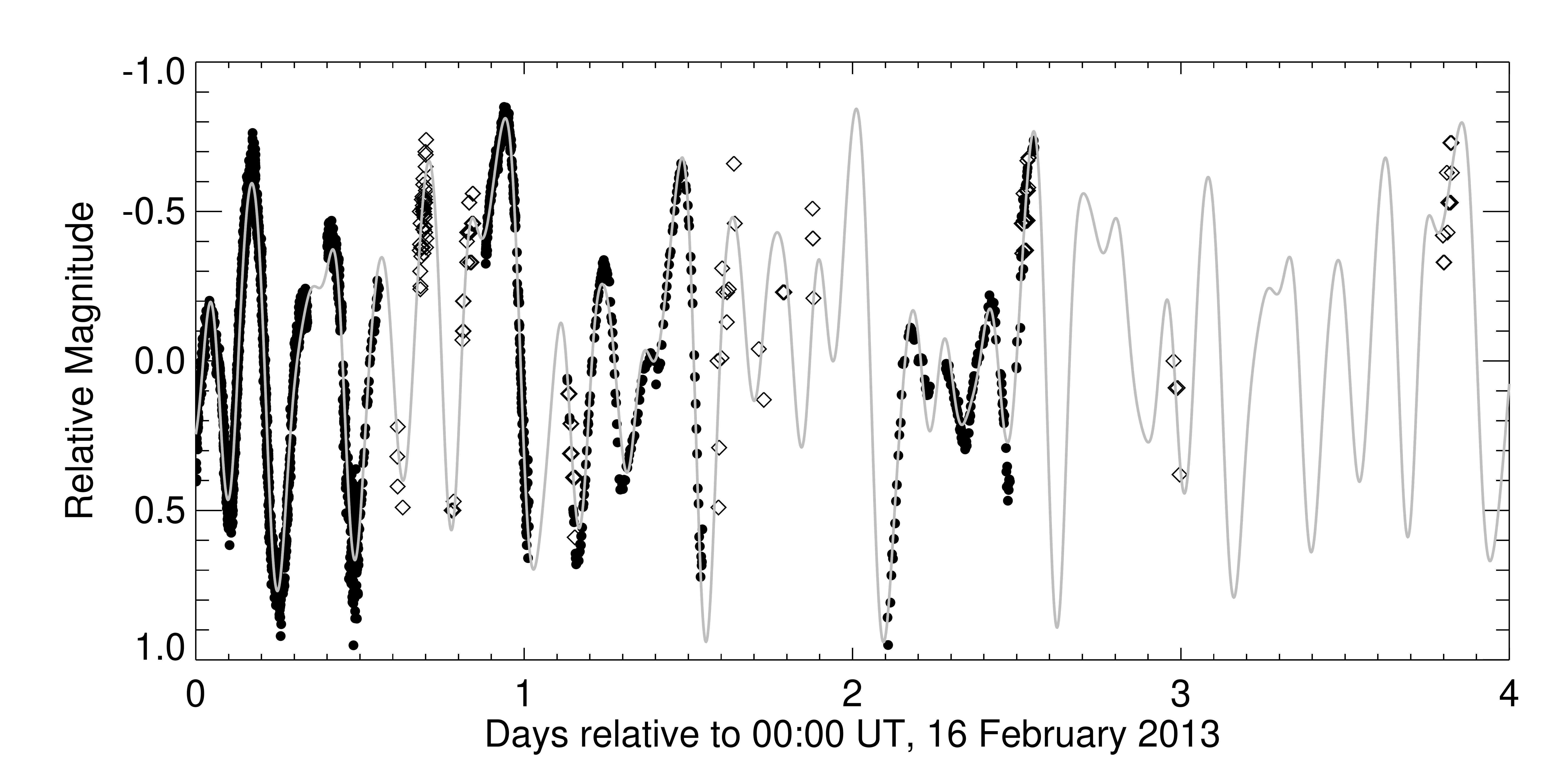}
\end{center}
\caption[]{Post-flyby rotational lightcurve of Duende clearly demonstrating a non-principal axis rotation state. The filled circles represent our observations; open diamonds represent data reported to the Minor Planet Center. The data are well fit with a dual period, tumbling asteroid model (grey curve), in this example with periods $P_1 = 8.71 \pm 0.03$ and $P_2 = 23.7 \pm 0.2$ hours. 
} 
\label{fig.post}
\end{figure}

The post-flyby lightcurve starts about 4.5 hours after closest approach, provides the strongest constraints on the rotational properties of Duende, and serves as a baseline for our analysis (Figure \ref{fig.post}). The more than 3000 individual data points are densely sampled with several segments overlapping in time, which provides a nice check on our magnitude calibrations. The non-repeating morphology of the post-flyby lightcurve is a clear indication of non-principal axis rotation (informally known as tumbling) and can be modeled with a two dimensional (dual period) Fourier series. This model follows the formalism of \citet{pravec05} and was applied to the combination of our observations and the selected data from the MPC. This procedure involved determining an error weighted least squares fit to the data with the IDL {\it mpfit} package using a second order 2D Fourier series \citep[Equation 6 in][]{pravec05} as the input function. This function includes 27 free parameters: an overall magnitude offset relative to zero, two distinct periods labeled $P_1$ and $P_2$, and 24 coefficients accounting for the amplitudes of each Fourier term. With the exception of the initial conditions for the two periods, all parameters in our model were initially set to zero. We also restricted the individual Fourier coefficients to $<0.5$ magnitude to limit net amplitude in the final functional fit. As noted in \citet{pravec05} and described in greater detail below the periods $P_1$ and $P_2$ can not always be assigned to specific rotational modes of the body due to issues of aliasing. We also tested 1st, 3rd and 4th order Fourier series which produced equivalent results. We settled on a second order function as a compromise between minimizing the number of free parameters in the model and providing a reasonable fit to the data.  Small vertical offsets within the signal-to-noise of each individual data set were manually applied to minimize the chi-squared residual of the model fits. Ultimately this approach produced strong constraints on the fitted periods, even though the fine-scale structure in the peaks and troughs of the lightcurve were not always perfectly fit. 

The criteria for using MPC data (Table \ref{tab.mpc}) in this analysis was as follows: we first determined a best-fit lightcurve to our observations only (filled dots in Figure \ref{fig.post}) while excluding the MPC data. The selected MPC data were then added in with observatory dependent magnitude offsets to account for the lack of precise photometric calibration that can be characteristic of MPC astrometric observations. These offsets were applied to bring the MPC data into agreement with the initial best-fit lightcurve model. We then refit the entire data set (ours + MPC) with errors of $\pm0.1$ magnitude assigned to the MPC points. The MPC data did not dramatically change the model results but did provide some temporal coverage at times where we were not able to obtain our own observations.

\begin{figure}[t]
\begin{center}
\includegraphics[width=14cm]{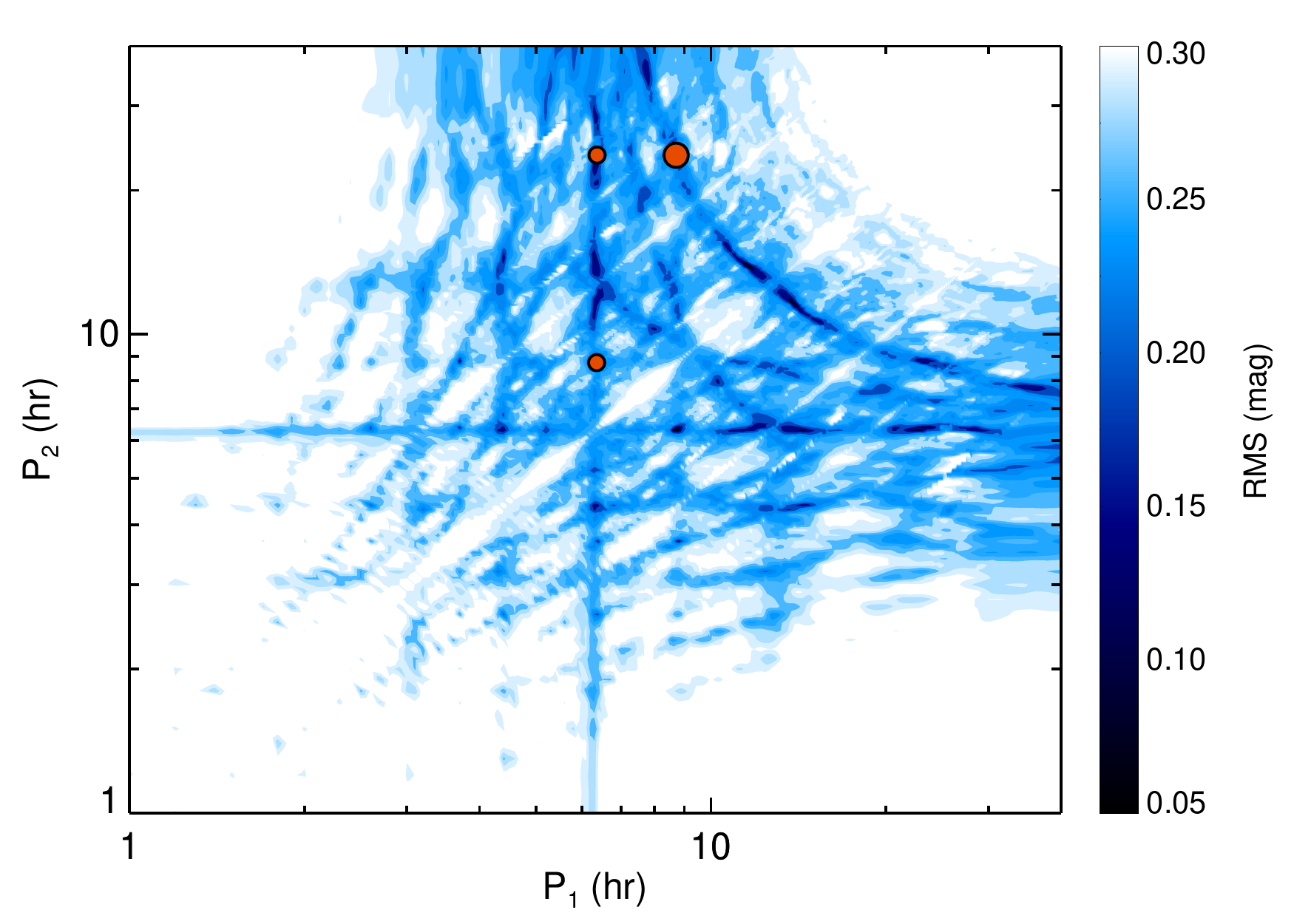}
\end{center}
\caption[]{Contours of RMS residuals of 2D Fourier series fits for a range of post-flyby $P_1$--$P_2$ period pairs. The RMS values are symmetric about the diagonal because the functional form of the Fourier series model is symmetric with regards to the two periods. The three lowest RMS solutions are marked with red dots, with the point for our preferred solution at $P_1=8.71$ hr and $P_2=23.7$ hr shown larger than the other two. These are the only solutions with RMS values below 0.1 mag.}
\label{fig.post_grid}
\end{figure}

To identify viable period solutions we performed a grid search across periods $P_1$ and $P_2$ from 1 to 40 hours in step sizes of 0.1 hours. At each grid point, the periods were held fixed while the Fourier coefficients were allowed to float, and an RMS residual (normalized by the number of data points) was computed for each period pair. A contour plot of those RMS residuals is shown in Figure \ref{fig.post_grid}. This analysis reveals the three lowest RMS solutions tabulated in Table \ref{tab.postSolutions}. The RMS residuals for these three solutions are sufficiently similar, i.e. the differences are much less than the noise in our data, that we do not find a strong preference for one period pair over the others. No other solutions between periods of 1-40 hours were found with RMS residuals less than 0.1 magnitude. Based on plausibility arguments presented here (\S\ref{sec.comparison}) and the complementary dynamical analysis of \citet{benson19}, we identify the period pair of 8.71 and 23.7 hours as the preferred interpretation of Duende's post-flyby spin state. The fit to the data shown in Figure \ref{fig.post} corresponds to these periods $P_1 = 8.71 \pm 0.03$ and $P_2 = 23.7 \pm 0.2$ hours, with errors bars determined based on the range of periods in which RMS residuals remained below 0.1 magnitude. 

\begin{table}[h]
\begin{tabular}{l|c|cccc}
\hline 
\hline
$P_1$, $P_2$	 	& RMS 	& $f_a$			& $f_b$				& $f_c$				& $f_d$ \\
(hr)	 			& (mag)	& (1.01 day$^{-1}$)	& (3.78 day$^{-1}$)		&(5.46 day$^{-1}$)		& (7.61 day$^{-1}$) \\
\hline
6.36, 8.73			& 0.067	& -				& 1/P$_1$				& 2/P$_2$				& 2/P$_1$	  \\
\bf 8.71, 23.7			& 0.079	& 1/P$_2$			& 1/P$_1$ + 1/P$_2$	& 2/P$_1$				& 2/P$_1$ + 2/P$_2$	  \\
6.37, 23.7			& 0.089	& 1/P$_2$			& 1/P$_1$				& 2/P$_1$ - 2/P$_2$		& 2/P$_1$	  \\
\hline
\hline
\end{tabular}
\caption{Top three period pair solutions to the post-flyby data. The corresponding frequency peaks in our FFT analysis are assigned to first and second order linear combinations of these periods. The preferred solution of 8.71 and 23.7 hours is highlighted.}
\label{tab.postSolutions}
\end{table}%

Previous analyses of Duende's post-flyby rotation state did not span a long enough temporal baseline to resolve complex rotation and/or were measured so near to close encounter that viewing aspect was highly variable which likely influenced the derived periods \citep{gary13,deleon13}. However, in both of these previous analyses a single period fit suggested a period around 9 hours, roughly consistent with the multi-hour periods in our more detailed analysis.

\begin{figure}[b]
\begin{center}
\includegraphics[width=14cm]{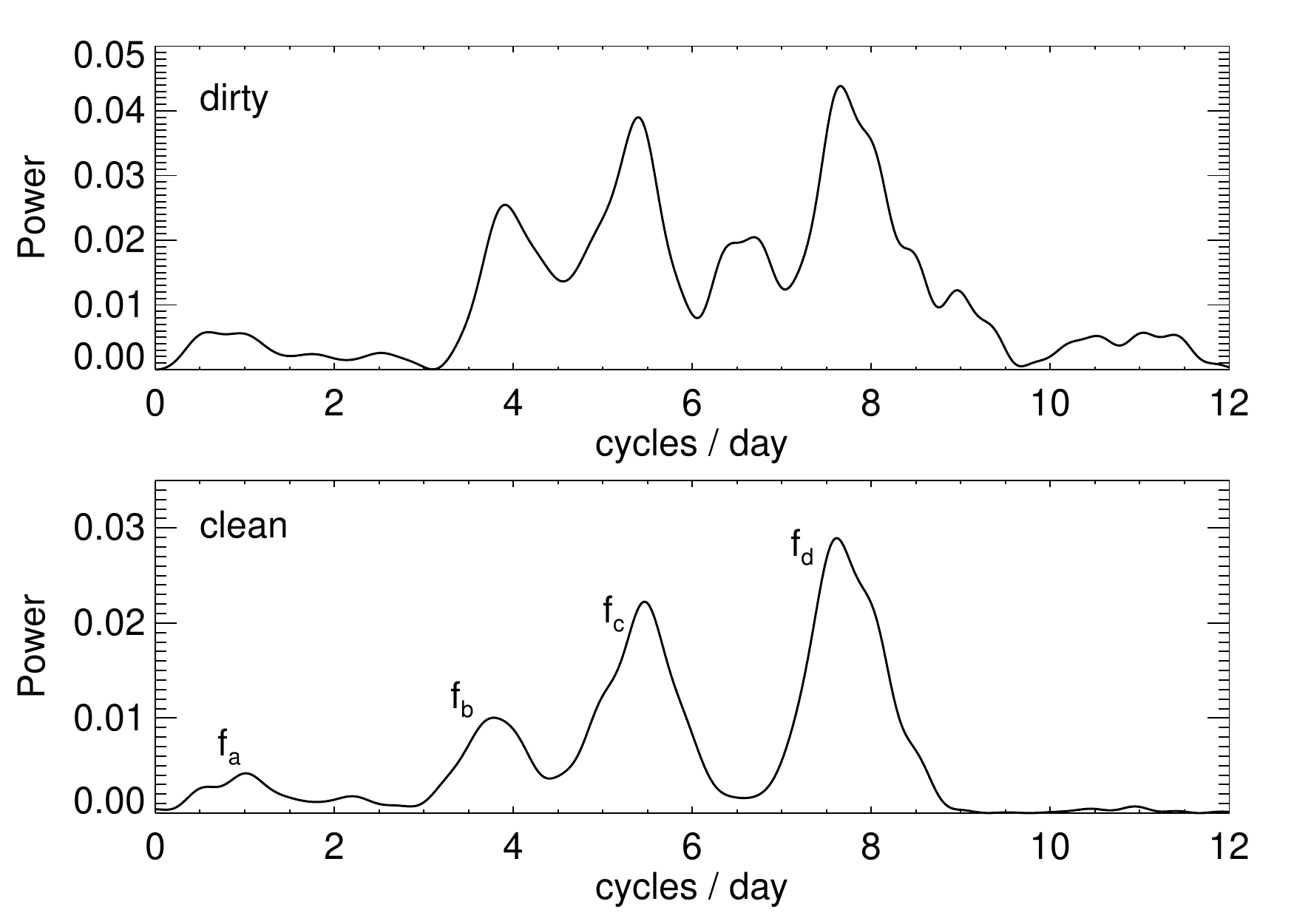}
\end{center}
\caption[]{FFT power spectra of post-flyby lightcurve. Both a clean (bottom) and dirty (top) version of the FFT are presented, see text for more details. The four highest peaks in the cleaned FFT are labeled in order of strength $f_{a-d}$ and can in most cases be directly mapped to the periods and linear combinations of those periods identified in our RMS fits to the data.
} 
\label{fig.fftPost}
\end{figure}

To complement our least-squares fitting of the post-flyby lightcurve, we also perform an FFT periodogram analysis of the data (Figure \ref{fig.fftPost}). This analysis involved computing a discrete Fourier transform of the lightcurve data to produce a ``dirty'' power spectrum, which includes aliases and other unwanted periodicities that, for example, may reflect the temporal sampling of the lightcurve based on observational cadences. This power spectrum was then ``cleaned'' using the WindowCLEAN algorithm, which numerically applies a deconvolution kernel to remove aliases in the power spectrum \citep{roberts87,belton88,mueller02}. The resulting clean spectrum in Figure \ref{fig.fftPost} clearly displays four prominent peaks that in most cases correlate well with the derived $P_1$ and $P_2$ from the RMS analysis. In the WindowCLEAN frequency space, peaks manifest at integer multiples or low order linear combinations of the periods present in the data. We assign the primary peaks in this FFT, labelled $f_{a-d}$, to specific periods and linear combinations of periods in Table \ref{tab.postSolutions}. For example, the two strongest peaks $f_c=5.46~{\rm day}^{-1}$ and $f_d=7.61~{\rm day}^{-1}$ closely correspond to the two periods in the minimum RMS solution, where $2/P_2 =$ 5.5 day$^{-1}$ and $2/P_1 =$ 7.6 day$^{-1}$. In this case the peak at $f_a=3.8~{\rm day}^{-1}$ is likely an alias at approximately 1/$P_1$. The presence of peaks at these specific aliases and linear combinations is fully consistent with analyses of lightcurves for other non-principal axis rotators \citep[e.g.][]{kryszczynska99,mueller02,samarasinha15}. In the dirty power spectrum, a prominent peak is seen around 6.5 cycles per day, which is then is removed by the cleaning algorithm. We interpret this as a combined alias of the peaks at 5.46 and 7.61 cycles per day. The bi-modal morphology of the dirty 6.5 cycle per day feature is consistent with this interpretation. Generally, major peaks in an FFT power spectrum will have symmetric aliases on either side of the primary frequency. Post-cleaning changes in the FFT power around 4, 6.5, and 9 cycles per day are consistent with such aliases connected to peaks c and d.

%
\subsection{Pre-Flyby and Discovery Epoch \label{sec.pre}}

Numerous issues related to weather and instrument failure resulted in less extensive lightcurve coverage during the discovery and pre-flyby epochs. As a result the lightcurve of Duende is not well constrained prior to the Earth encounter in February 2013.

\begin{figure}[b]
\begin{center}
\includegraphics[width=14cm]{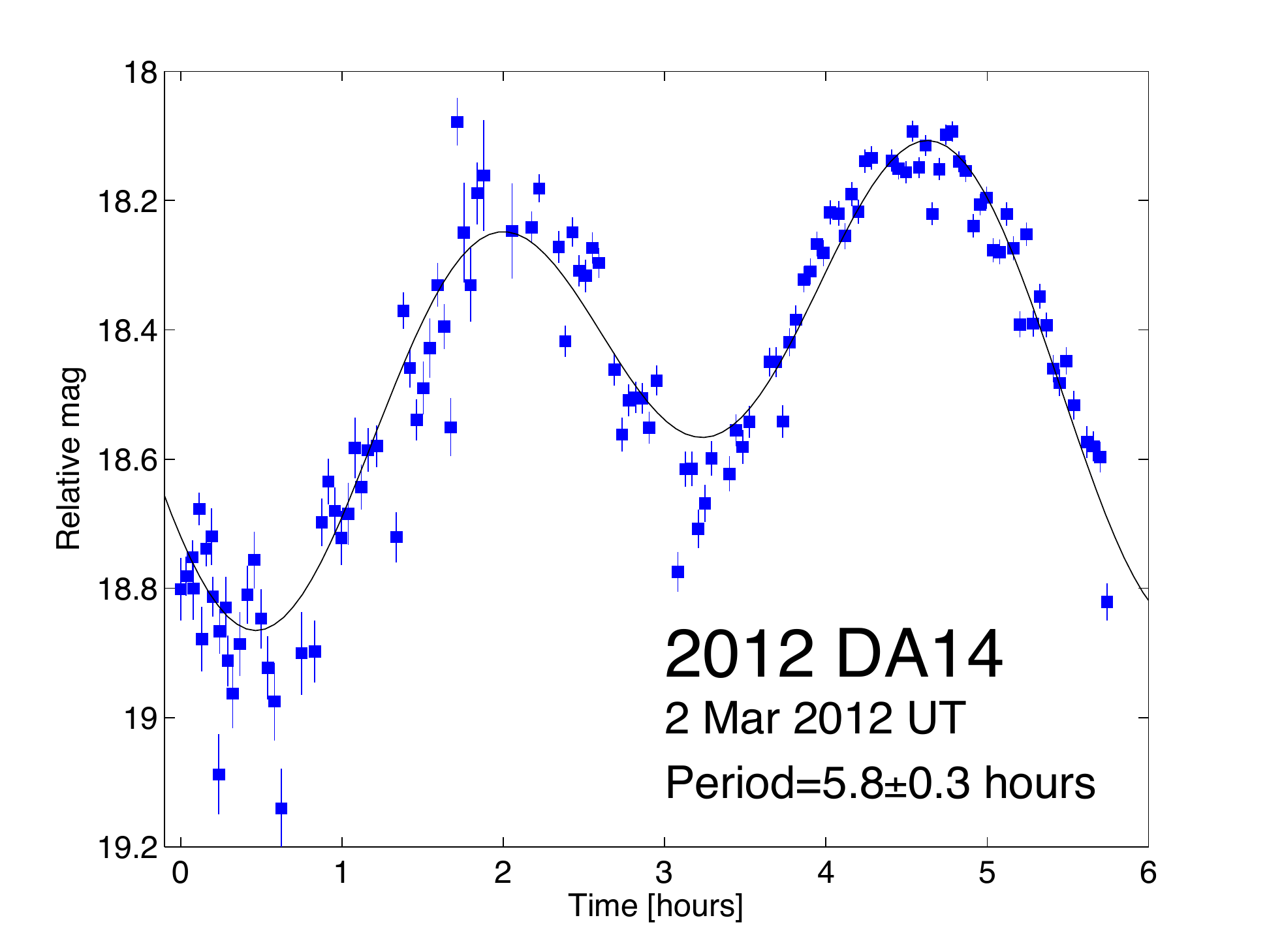}
\end{center}
\caption[]{Pre-flyby lightcurve of Duende taken during the discovery epoch in March of 2012. A single-period fit suggests a rotation period of approximately 5.8 hours. However, this is not an accurate representation of an object in non-principal axis rotation.} 
\label{fig.disclc}
\end{figure}

Data were acquired with the SNIFS instrument at the University of Hawaii 2.2m telescope on Mauna Kea within a week of Duende's discovery (Figure \ref{fig.disclc}). These data did not span a sufficiently long baseline to resolve complex rotation, though a single period solution suggests a period of $5.8 \pm 0.3$ hours. However, a single period solution over such a short time interval for an object in complex rotation is undoubtedly misleading \citep{pravec05}. 

\begin{figure}[]
\begin{center}
\includegraphics[width=14cm]{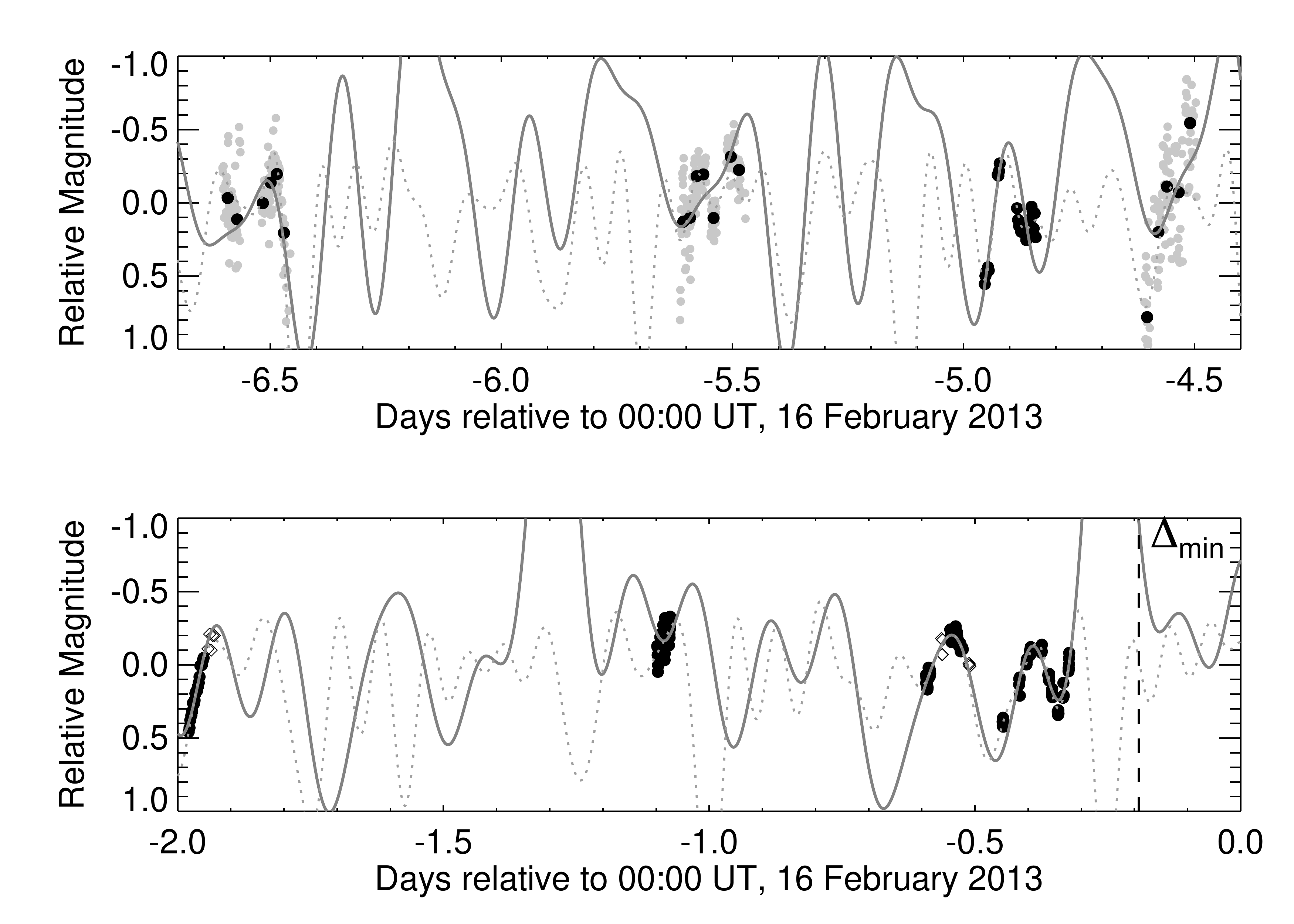}
\end{center}
\caption[]{Pre-flyby rotational lightcurve of Duende. The time of closest approach is labeled as $\Delta_{min}$ (dashed line). The filled black circles represent our observations; the open diamonds represent data reported to the Minor Planet Center. Low S/N data (grey dots) have been combined to produce time-averaged photometric points. The large gaps in rotational coverage make it difficult to uniquely constrain the lightcurve in the same manner as the post-flyby data. Dual period fits to the data are shown for the lowest RMS solution $P'_1 = 4.5$ hours and $P'_2 = 16.4$ hours (dotted), and our preferred solution of $P'_1 = 8.37$ hours and $P'_2 = 24.18$ hours (solid grey).} 
\label{fig.prelc}
\end{figure}

Pre-flyby observations in February 2013 were conducted from a number of observatories in the southern hemisphere (Table \ref{tab.phot}). These observations intermittently sampled a time span starting approximately 6.5 days before closest approach (Figure \ref{fig.prelc}). We show in this figure representative fits to the data using the same dual period Fourier model \citep{pravec05} with periods $P'_1 = 4.5$ hours and $P'_2 = 16.4$ hours, our lowest RMS fit, and the preferred period paring of $P'_1 = 8.37$ hours and $P'_2 = 24.18$ hours (\S\ref{sec.comparison}). Data from individual observatories have been iteratively offset to minimize the residual of the fit. The three data sets from Mt. John Observatory, around days -6.5, -5.5 and -4.5, are about an order of magnitude noisier than our other data sets and thus have been temporally re-binned to an interval of $\sim$0.02 hours. For fitting purposes, the errors for these time averaged points were set to the standard deviation of the binned data.

\begin{figure}[]
\begin{center}
\includegraphics[width=14cm]{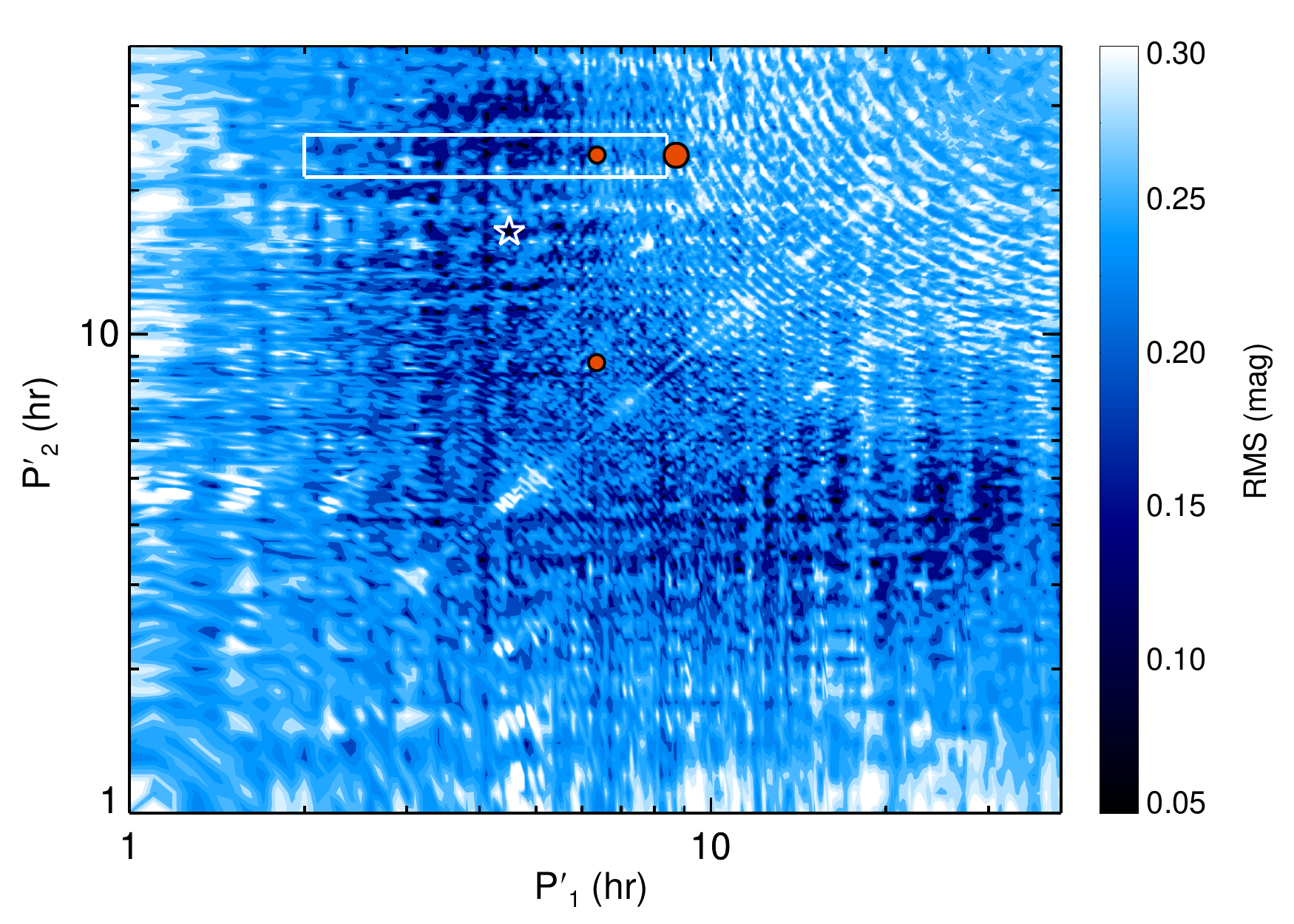}
\end{center}
\caption[]{Contours of RMS residuals of 2D Fourier series fits for a range of pre-flyby $P'_1$--$P'_2$ period pairs. The preferred post-flyby solution of $P_1 = 8.71$ and $P_2 = 23.7$ hours is shown as the large red dot and falls in a region where there are no low-RMS pre-flyby solutions. The smaller red dots are the other two post-flyby solutions (Table \ref{tab.postSolutions}). Though there is no single combination of periods that provide a unique solution to the pre-flyby data, there are a large number of local minima that could be valid. The global minimum is indicated with the white star. The white box centered around $P'_2 = 23.7$ encloses a set of low RMS solutions that represent our preferred interpretation based on dynamical arguments \citep{benson19}.}
\label{fig.grid}
\end{figure}

To analyze the pre-flyby data, we employed the same 2D Fourier series function and the same 0.1-hr period grid search as was done for the post-flyby data (\S\ref{sec.post}). The result of this grid search is shown in Figure \ref{fig.grid} with the contours shown on the same color bar. Even though this figure shows that there is no clear single solution to the pre-flyby data, there are a number of key aspects that provide clues about the pre-flyby spin state. First, the consistently high RMS values along the diagonal (and at integer multiples of the diagonal) is a clear indication that Duende was in non-principal axis rotation prior to the encounter, i.e. two periods are needed to fit the pre-flyby data. Second, the RMS contours are not randomly distributed, but instead show clear structure. Of the 160,801 period combinations presented in this figure, 778 of them have RMS residuals less than 0.1 magnitude, where the lowest at $P'_1 = 4.5$ and $P'_2 = 16.4$ hr has an RMS = 0.059. This large number of local minima do appear in several well-defined horizontal and vertical bands (reflected symmetrically about the diagonal) with prominent examples centered around 4.1, 6.2, 8.3, 12.5, 16.5, and 24.6 hours. The width of these bands and many of the local minima are greater than the grid spacing of 0.1 hours, thus suggesting that these features are a representation of real periodicity and are not dominated by fluctuations induced by noise in the data.  However, this does suggest that the combination of noise in the data and grid spacing leads to pre-flyby period solutions that are precise to no better than $\sim0.1$ hour. 

In general, for the pre-flyby case, we find that a number of solutions are more-or-less statistically indistinguishable ($\sim0.01$ magnitude difference in residuals) with RMS values approaching the noise level of our data ($\sim0.1$ magnitudes). Several of these solutions are consistent with the peaks in a WindowClean Fourier analysis presented below. In addition, a number of the period pairs identified in this RMS analysis are consistent with the allowable spin states expected from a detailed analysis of Duende's solid body dynamics \citep{benson19}. The results of this detailed analysis are leveraged in \S\ref{sec.comparison} to identify a single set of preferred pre-flyby periods for comparison to the post-flyby spin state.

Other than limiting the Fourier coefficients to $<0.5$ magnitudes in our fitting algorithm (\S\ref{sec.post}), the net amplitudes of the pre-flyby fits were not restricted in any way. However, due to the large temporal gaps in the pre-flyby data, this resulted in some cases where the model amplitudes were much greater than the well-defined post-flyby amplitude of 1.9 magnitudes (Figure \ref{fig.post}). Without detailed knowledge of Duende's shape, which we don't currently have, we can not make a clear estimate for the expected amplitude at the time of the pre-flyby viewing geometry. Instead, based on the ensemble of measured amplitudes for other similarly sized NEOs, including those in non-principal axis rotation \citep{warner09}, we expect Duende to have a maximum lightcurve amplitude of less than about 2.5 magnitudes. In our pre-flyby grid search we find that 70\% of fits have amplitudes $\leq2.5$ magnitudes, suggesting that the majority of fitted amplitudes are reasonable. The exact threshold for when a fitted amplitude becomes unphysical remains unknown.

\begin{figure}[h!]
\begin{center}
\includegraphics[width=14cm]{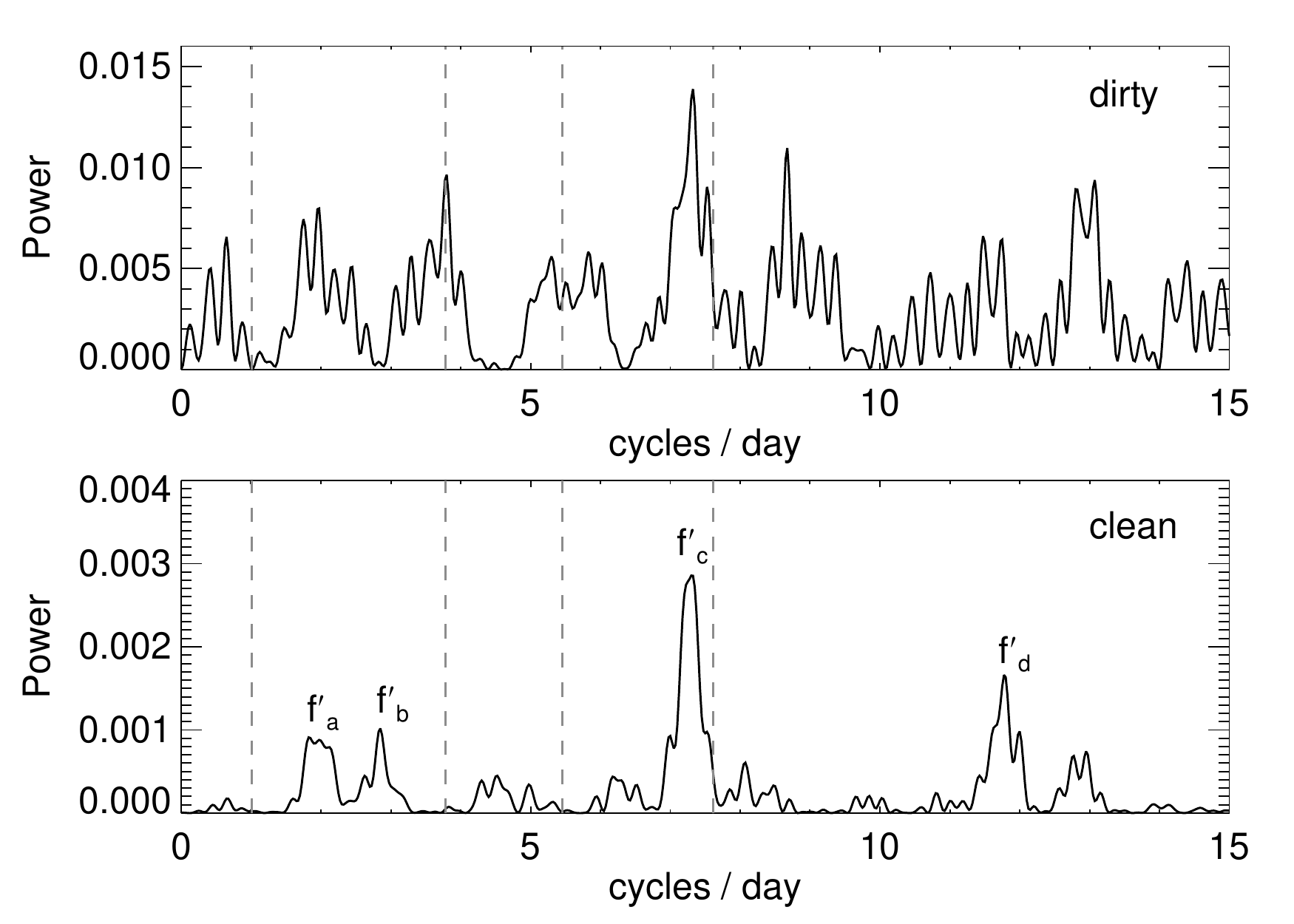}
\end{center}
\caption[]{FFT power spectra of the pre-flyby lightcurve data. The locations of the 4 most prominent peaks in the post-flyby FFT are shown as dashed grey lines. The cleaned power spectrum (bottom) shows several peaks ($f^\prime_{a-d}$) which are discussed in the text. Several of these can be mapped to periods identified in our RMS fitting analysis. In general the peaks here do not match those in the post-flyby FFT.
} 
\label{fig.fftPre}
\end{figure}

To provide a complementary probe of periodic signatures in the pre-flyby data, the same FFT WindowClean analysis from \S\ref{sec.post} was applied to these data (Figure \ref{fig.fftPre}). Though the sparseness and low quality of the pre-flyby data results in a much noisier FFT, this result does display different peaks than the post-flyby case. With the understanding that additional physically meaningful peaks could be obscured by noise, there are useful comparisons between these FFT peaks and the periods identified in the RMS fitting analysis. The most prominent peaks in Figure \ref{fig.fftPre} are at frequencies around $f^\prime_a$ = 2.0, $f^\prime_b$ = 2.9, $f^\prime_c$ = 7.2, and $f^\prime_d$ = 11.7 cycles per day (Table \ref{tab.preSolutions}). The second highest peak around 11.7 cycles per day is a possible counterpart to the series of $\sim4.1$ hour solutions in the RMS fits (2 / 4.1 hr = 11.7 day$^{-1}$). The largest peak in this FFT occurs around 7.2 cycles per day. This is close to the peak associated with the second order linear combination of our preferred pre-flyby period pair of $P^\prime_1=8.37$ and $P^\prime_2=24.2$ (\S\ref{sec.comparison}). It also is close to $2/P^\prime_1$ for period solutions around 6 hours. The peak at 2.9 cycles per day is close to $1/P^\prime_1$ for periods around 8 hours. The peak centered on 2 cycles per day may correspond to $2/P^\prime_2$ for solutions around 24 hours.

\begin{table}[h]
\begin{tabular}{l | l | cccc}
\hline 
\hline
$P'_1$, $P'_2$ 		& RMS 	& $f^\prime_a$ 		& $f^\prime_b$ 		& $f^\prime_c$ 		& $f^ \prime_d$ \\
(hr) 				& (mag)	& (2.0 day$^{-1}$)	& (2.9 day$^{-1}$)	& (7.2 day$^{-1}$)	& (11.7 day$^{-1}$) \\
\hline
4.5, 16.4			& 0.06	& - 				& 2/P$^\prime_2$ 	& 1/P$^\prime_1$ + 1/P$^\prime_2$ 	& 2/P$^\prime_1$ \\
6.26, 24.0		& 0.09	& 2/P$^\prime_2$ 	& 1/P$^\prime_1$ - 1/P$^\prime_2$ 		& 2/P$^\prime_1$ 	& 3/P$^\prime_1$ \\
6.37, 22.8		& 0.09	& 2/P$^\prime_2$ 	& 1/P$^\prime_1$ - 1/P$^\prime_2$ 		& 2/P$^\prime_1$ 	& - \\
6.84, 24.0		& 0.10	& 2/P$^\prime_2$ 	& -	& 2/P$^\prime_1$ 	& -  \\
7.56, 23.9		& 0.10	& 2/P$^\prime_2$ 	& 1/P$^\prime_1$  		& - 	& - \\
\bf 8.37, \bf 24.2	& 0.10	& 2/P$^\prime_2$ 	& 1/P$^\prime_1$ 		& 2/P$^\prime_1$ + 2/P$^\prime_2$ 	& 4/P$^\prime_1$ \\
\hline
\hline
\end{tabular}
\caption{Selection of pre-flyby period solutions. Our preferred solution is highlighted.}
\label{tab.preSolutions}
\label{lasttable}
\end{table}%

In general the FFT for the pre-flyby data displays different peaks than that of the post-flyby case (Figure \ref{fig.fftPost}). However, there are some similarities, particularly for our preferred period solutions. Both pre and post-flyby power spectra show peaks associated with periods around 24 hours (2/24.2 hr for the former, 1/23.7 hr for the latter), and with periods around 8 hours (1/8.37 hr for the former, 1/8.71 hr for the latter). The key question that remains is whether these differences, including the different best-fit periods in our RMS analysis, are due to noise in the pre-flyby data, are due to changes in observing geometry pre- and post-flyby, and/or  indicate a change in rotation state due to tidal effects during the encounter. We address this question in the following section, noting that the more comprehensive analysis of the solid body dynamics presented in \citet{benson19} provides results that complement arguments presented here. We note here an important point that generalized models for bodies in NPA rotation find that viewing geometry changes can modify the strength of FFT peaks, but typically do not cause a shift of the peaks in frequency space \citep{samarasinha15}.

%
%
\section{Pre/Post-flyby Comparison \label{sec.comparison}}

Our initial objective for the lightcurve investigation was to probe for any detectable changes in the rotation state of the asteroid. While the pre-flyby data are not of sufficient quality to definitively address this goal, we do compare analyses of the pre and post-flyby data to identify preferred pre-flyby periods and to address the plausibility that Duende experienced rotational changes during the planetary encounter. We investigate in detail the low RMS solutions returned in our least squares analysis (\S\ref{sec.compRMS}) and the implications of our FFT spectral analysis (\S\ref{sec.compFFT}).

\begin{figure}[t!]
\begin{center}
\includegraphics[width=14cm]{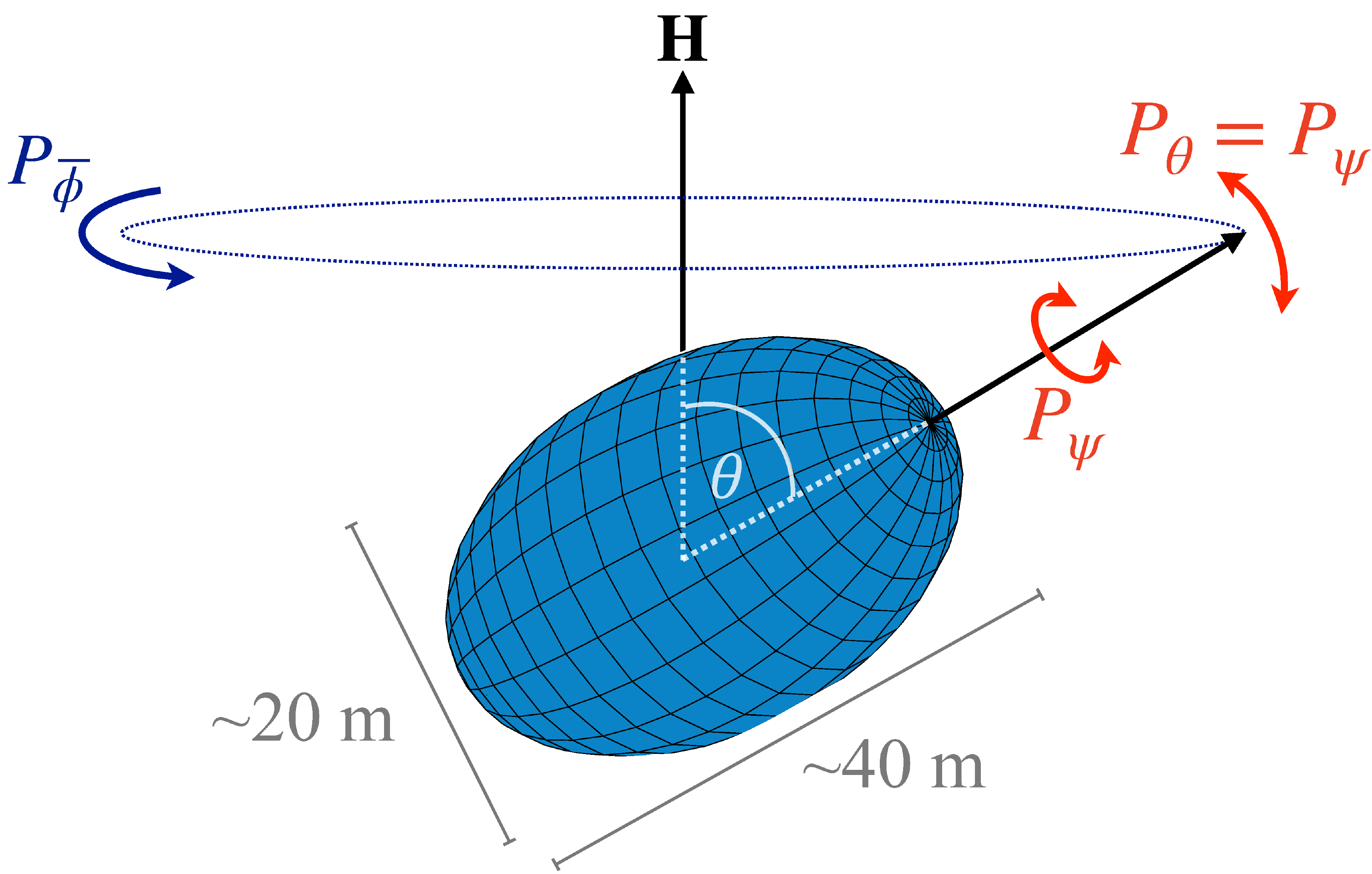}
\end{center}
\caption[]{Rotational dynamics of an ellipsoid in non-principal, short-axis mode (SAM) rotation.  Duende is roughly a prolate ellipsoid 20 meters along its short axes and 40 meters along its major axis. SAM rotation can be described by oscillation of a body around its axis of symmetry ($\psi$) and rotation of the symmetry axis about the angular momentum vector ($\vec{H}$). The rotation periods associated with these two axes are $P_\psi$ and $P_{\bar{\phi}}$ respectively. Based on constraints from the radar observations and the solid body analysis in \citet{benson19}, the nutation angle $\theta$ is expected to be large, $\sim70-85^\circ$. For axis convention shown here, the nodding rate $P_\theta=P_\psi$. 
} 
\label{fig.geo}
\end{figure}

A brief overview of short-axis non-principal axis rotation facilitates the following discussion (Figure \ref{fig.geo}). More detailed descriptions of the SAM coordinate system, rotation mode, and justification for its relevance to Duende, are provided in \citet{benson19}. A SAM state can be described by rotation about several axes. The long axis of the body or the symmetry axis ($\psi$) rotates with an average angular precessionary period $P_{\bar{\phi}}$ about the net angular momentum vector $\vec{H}$. The symmetry axis also nods with a period $P_\theta$ around the nutation angle $\theta$. For for the long axis convention shown here, $P_\theta=P_\psi$ \citep{benson19}. Lastly, the body oscillates around the symmetry axis with angular amplitude less than $90^\circ$ and a period equal to $P_\psi$. This description is based on \citet{greenwood87} and follows the notation of \citet{pravec05}. 

\subsection{RMS Fitting \label{sec.compRMS}}

Attempts to fit the pre-flyby data with periods equal to those measured post-flyby resulted in RMS residuals that were a factor of at least two higher than the best fit RMS values of $\sim0.06$ magnitudes (Table \ref{tab.preSolutions}). Forcing fits to the pre-flyby data with the post-flyby periods in Table \ref{tab.postSolutions} resulted in RMS values of 0.12 for $P'_1 = 6.36$ hr and $P'_2 = 8.73$ hr, of 0.19 for $P'_1 = 8.71$ hr and $P'_2 = 23.7$ hr, and 0.12 for $P'_1 = 6.37$ hr and $P'_2 = 23.7$ hr. With mean residuals this high ($>0.1$ magnitude) there are clear discrepancies between the data and the fit, in some cases with the fits not tracking clear trends (e.g. increasing or decreasing magnitude) in the data.

Based on an analysis of the solid body dynamics \citep{benson19}, the post-flyby period pair with the second lowest RMS, $P_1 = 8.71$ hr and $P_2 = 23.7$ (Table \ref{tab.postSolutions}), offers the most plausible description of the post-flyby rotation state. Furthermore, this dynamical analysis suggests that $P_1 = P_\phi$ and $P_2 = P_\psi$. We expect pre-flyby solutions to display similar values of $P_\psi$ due to the difficulty in tidally inducing period changes about the symmetry axis. Considering the uncertainties associated with the pre-flyby data, we suggest that pre-flyby periods within $\sim10\%$ of $P_\psi=$23.7 hours would be broadly consistent with a constant $P_\psi$ throughout the flyby. With this as a rough guideline we identify a number of $P^\prime_1$ and $P^\prime_2$ period solutions -- enclosed by the white box in Figure \ref{fig.grid} -- for which a low RMS is realized.  This box encloses a range of $P'_1$ values from about 2 hours up to 8.4 hours, and $P'_2$ values from about 21 to 26 hours. Although these low RMS solutions go down to $P'_1$ as small as 2 hours, a suite of numerical models looking at the outcome of randomized encounter geometries find that the smallest possible pre-flyby $P_{\bar{\phi}}$ is about 6 hours assuming a post-flyby value of 8.71 hr \citep[Figure 10 in][]{benson19}. We thus set a  boundary at $P'_1=6$ hours, anticipating more modest period changes in line with the most likely simulated outcomes \citep{benson19}, and then search for best fit solutions within the ranges $6 < P'_1 < 8.4$ hours and $21 < P'_2 < 26$ hours without restricting the periods as was done in the grid search. This leads to a set of 5  pre-flyby solutions that are close to the post-flyby periods and have $P'_2$ within 10\% of 23.7 hours (Table \ref{tab.preSolutions}). The RMS residuals for these solutions are all equal to about 0.1 magnitude, which is not the lowest found in the global grid search (Figure \ref{fig.grid}), but we favor these solutions as most plausible based on the solid body dynamics \citep{benson19} and based on the FFT analysis below. Due to the sparseness and lower quality of the pre-flyby data it is unclear whether the difference in RMS between these five solutions ($\sim0.1$) and the global minimum (=0.06) is statistically significant. Nevertheless, it is interesting to consider these solutions in slightly more detail. All five of these pre-flyby solutions centered around $P'_2=23.7$ hr are at significantly smaller $P'_1$ than the preferred post flyby $P_1 = 8.7$ hours. If we thus interpret these in the same manner as the post-flyby periods, then this suggests the intriguing possibility that Duende's rotation slowed down, i.e. $P_\phi$ increased during the planetary flyby. The specific pre-flyby solution of $P'_1=8.37$ hr and $P'_2=24.18$ hr (Table \ref{tab.preSolutions}) falls nearest the most likely outcome of these numerical models. At first glance, it would seem that the pre-flyby solution of 6.37 and 22.78 hours (Table \ref{tab.preSolutions}) could have gone unchanged to the post-flyby solution of 6.37 and 23.7 hours (Table \ref{tab.postSolutions}). However, this scenario seems inconsistent with the delay-Doppler radar constraint requiring $P_{\bar{\phi}}\sim8$ hours. As such, we suggest that the RMS fits to the data suggest a small increase in $P_{\bar{\phi}}$ from 8.37 hr pre-flyby to 8.71 hr post-flyby.

\subsection{FFT Power Spectra \label{sec.compFFT}}

We also employ the FFT analysis as a means to further probe the pre and post-flyby rotation states. The FFT power spectra alone (Figures \ref{fig.fftPost} and \ref{fig.fftPre}) unfortunately do not provide conclusive evidence for a change in rotation state. The single strongest peaks in each of these FFTs are roughly in the same location: 7.6 cycles per day for the post-flyby case and 7.2 cycles per day for the pre-flyby case. For our preferred post-flyby solution of $P_{\bar{\phi}} = 8.71$ hr and $P_\psi = 23.7$ hr, these would correspond to the second order linear combination 2/$P_{\bar{\phi}}$ + 2/P$_\psi$. For these periods the pre-flyby FFT peaks around 2 and 2.9 could correspond to 2/$P_\psi$ and 1/$P_{\bar{\phi}}$ respectively. However, the complete absence of the prominent post-flyby peaks at 3.78 and 5.46 cycles per day in Figure \ref{fig.fftPre} suggests that the pre-flyby data either sample a different rotation state or are too sparse to efficiently probe all of the frequencies present in the post-flyby FFT. It is expected that if Duende's rotation state did not change, then the frequencies of the FFT peaks should remain fixed while the relative intensities of the peaks might change \citep{samarasinha15}.

To address whether these differences between the two FFTs are significant, we run a simple Monte Carlo experiment to re-sample the post-flyby lightcurve with the sparse cadence of the pre-flyby observations. This test will address two specific objectives: (1) determine whether such sparse sampling can accurately reproduce the major peaks retrieved from denser observations, and (2) determine the likelihood of whether the different peaks in the noisy pre-flyby FFT could simply be due to sparse sampling of an unchanged post-flyby rotation state. We run this test on the model fit to the post-flyby data so as to have a noise-free representation of the lightcurve of a non-principal axis rotator. This model is extrapolated out for 7 days relative to time zero defined by the start of the post-flyby data set. We repeatedly sample this ideal lightcurve at specific times that exactly mimic the pre-flyby observations. For each sample set we randomly assign from a uniform distribution a 0 to 0.5 day offset relative to time zero, and we assign from a gaussian distribution with a full-width-half-maximum = 0.1 magnitude random noise to each sample point. For each sample set we run the WindowClean analysis (\S\ref{sec.post}) and record the five peaks with the highest power in the cleaned FFT. This sampling procedure is repeated 1000 times. 

\begin{figure}[h!]
\begin{center}
\includegraphics[width=14cm]{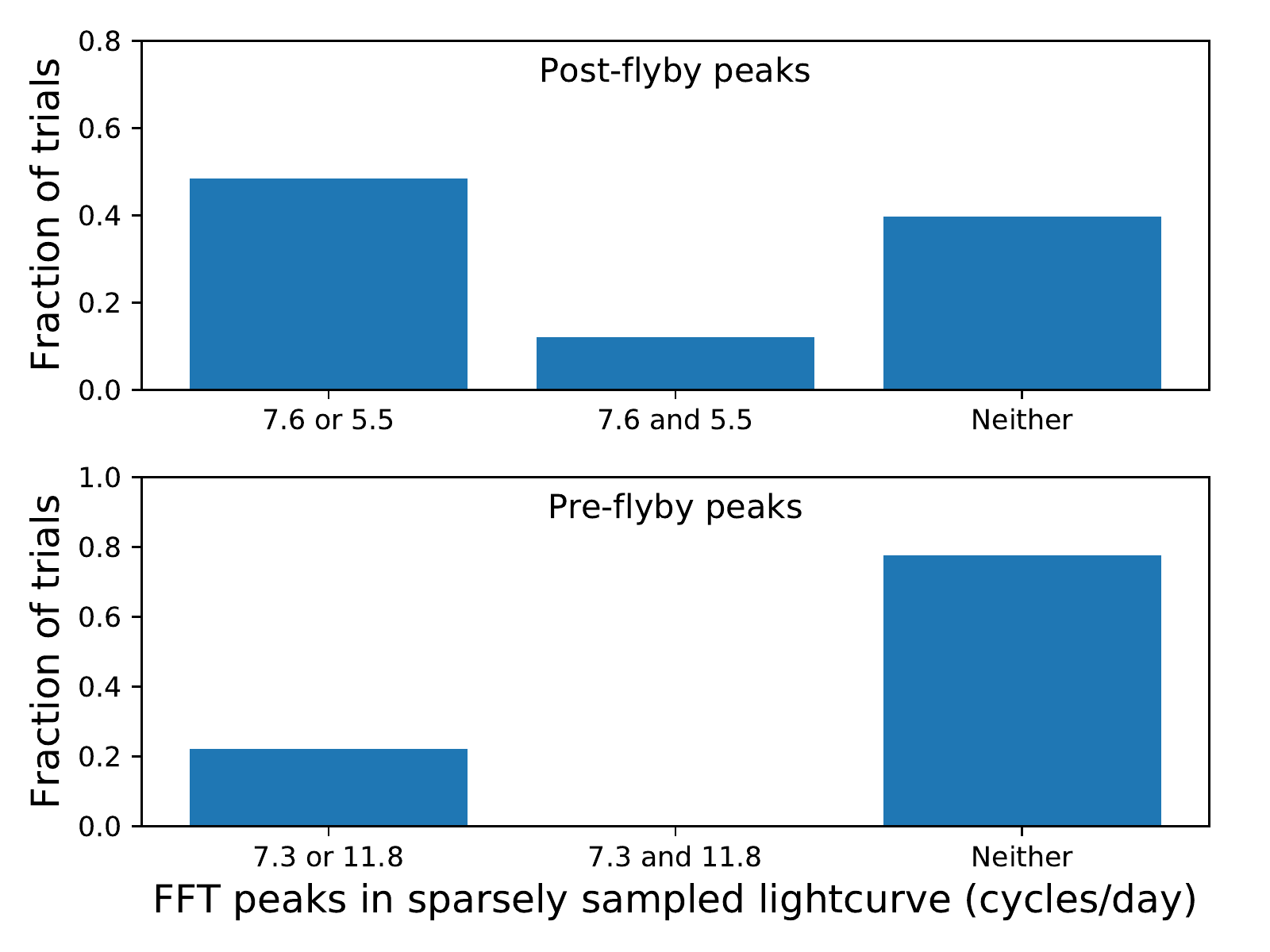}
\end{center}
\caption[]{Fraction of 1000 Monte Carlo trials in which the specified FFT peaks are retrieved when sparsely sampling a model of the post-flyby lightcurve at a cadence that mimics the pre-flyby observations. The top panel represents the ability of the sparse sampling scheme to retrieve the ``correct" solution, in this case the 7.6 cycle/day and/or the 5.5 cycle/day primary peaks from the post-flyby FFT. The sparse sampling retrieves one of these peaks in 48\% of trials, both of them in 12\% of trials, and neither in 40\% of trials. The bottom panel represents the likelihood that the primary peaks seen in the pre-flyby FFT are an artifact of sparse sampling. This shows it is unlikely that sparse sampling alone could result in an FFT with peaks at 7.3 and 11.8 cycles per day, thereby suggesting that a change in rotation state may have occurred.
} 
\label{fig.resample}
\label{lastfig}
\end{figure}

To address the first objective, we first consider the fraction of the 1000 trials that reproduce the ``true" FFT peaks. Specifically we look for whether the sparse sampling can retrieve the two major post-flyby FFT peaks at 7.6 and 5.5 cycles per day (Figure \ref{fig.resample}). We consider a peak to match one of these frequencies if it comes within 0.15 cycles / day, roughly the full-width-half-maximum of the peaks in the pre-flyby FFT power spectrum (Figure \ref{fig.fftPre}). We find that at least one of the these peaks is retrieved in 48\% of trials, that both of them are present in 12\% of trials, and that neither of them shows up in 40\% of trials. Therefore, in a slight majority (60\%) of cases the sparse sampling does retrieve accurate information about at least one of the frequencies present in the underlying lightcurve. This suggests that the pre-flyby FFT is more likely that not to contain physically meaningful information about rotation state.

Since it is suggestive that the sparse sampling can retrieve physically meaningful frequencies, we now address the likelihood that the peaks in the pre-flyby FFT could be a spurious outcome of noise and under-sampling. The two largest peaks in the pre-flyby FFT are at 7.3 and 11.8 cycles per day (Figure \ref{fig.fftPre}). If the rotation state went unchanged during the flyby, then we would expect sparse sampling of the post-flyby model to retrieve at least one of those peaks in a majority ($\sim60\%$) of trials. This is not the case (Figure \ref{fig.resample}). In nearly 80\% of our trials neither the 7.3 nor the 11.8 peak were detected. Again a threshold of 0.15 cycles / day was used to determine whether a trial peak matched the 7.3 and 11.8 pre-flyby frequencies. In only 1 out of the 1000 trials did the sparse sampling of the post-flyby model find peaks at both 7.3 and 11.8 cycles per day.

The combination of these results suggests that if the rotation state of Duende went unchanged, then the sparse sampling during the pre-flyby epoch was more likely than not to retrieve within our threshold of 0.15 cycles per day at least one of the major FFT peaks seen in the densely observed post-flyby case. Of course the marked viewing geometry changes associated with our observations (Figure \ref{fig.obs_geo}) are a likely cause of different lightcurve morphologies pre and post-flyby. However, models suggest that such changes would affect the relative strength of FFT peaks and not cause peaks to shift significantly in frequency space \citep{samarasinha15}. Thus the differences between the pre and post flyby FFT power spectra are unlikely to be attributable to sparse observational sampling or viewing geometry changes. If in fact these differences are physically meaningful, then this offers an alternative analysis to the RMS fitting that also suggests Duende may have experienced rotational changes during the flyby.

%
%
\section{Summary and Discussion \label{sec.disc}}

We have presented data from an extensive observing campaign of visible wavelength spectroscopy and photometry focused on the near-Earth asteroid 367943 Duende. Observations were conducted at the time of this object's discovery in 2012, and surrounding its near-Earth flyby at a distance of 4.2 Earth radii on 15 February 2013. In general the observations were focused on monitoring the object through the flyby to search for evidence of physical changes induced by gravitational interactions with the Earth. The spectroscopic observations suggest an L-, K-, or Ld- spectral type, but did not reveal any spectral changes within the systematic uncertainty of the data. Lightcurve photometry revealed that Duende was clearly in a non-principal axis rotation state both prior to and following the planetary encounter. The multi-hour rotation rate makes Duende the slowest known non-principal axis rotator for asteroids smaller than 100 meters. Such an excited rotation state could have been induced during prior planetary encounters \citep{scheeres00}. For an object with such a low Earth encounter distance, it is unlikely that the 2013 flyby was the first time it experienced close passage to the Earth.

It is not always possible to assign the rotation rates retrieved from period analysis techniques to physical rotation axes of a body. However, the thorough coverage of our post-flyby lightcurve coupled with delay-doppler radar imaging and inertial constraints \citep{benson19} strongly suggests that Duende is now in a non-principal, short-axis mode rotation state with a body-rotation period $P_\psi=23.67$ hr and a precession period $P_{\bar{\phi}}=8.71$ hr. Dynamical considerations by \citet{benson19} suggest that any changes in rotation state would have been largest around the precession axis $\phi$. Simulations performed by these authors suggest that Duende could have experienced a wide range of possible changes in rotation state during the flyby (e.g. $P_{\bar{\phi}}$ changing anywhere from -25\% up to +200\%), with the most likely outcome a more modest percent-level increase in $P_{\bar{\phi}}$.

Our pre-flyby lightcurve was much less extensive and resulted in looser constraints on the pre-encounter rotation state. Leveraging what we learned from the post-flyby data, we provide a series of plausibility arguments that suggests a spin down scenario is consistent with the data. Our preferred solution based on fits to the data, Fourier analyses, and dynamical arguments suggests that $P_{\bar{\phi}}$ changed from 8.4 to 8.7 hours and that $P_\psi$ remained roughly unchanged at a rate around 24 hours. This result is consistent with that of \citet{benson19} where more detailed dynamics and generation of synthetic lightcurves were considered. However, the sparse temporal coverage and lower signal-to-noise of the pre-flyby photometry ultimately made it difficult to definitively determine whether a spin down did in fact happen.

The results from this campaign highlight the need for extensive data coverage on both sides of planetary encounter. The slow multi-hour and non-principal axis rotation state of Duende made it particularly challenging to collect sufficient lightcurve photometry, though such a rotation state, in comparison to a fast principal axis rotator, would enhance any gravitationally induced effects during planetary encounters \citep{scheeres00}. Our analyses of the lightcurves suggest that the analytic tools we employed (functional fitting, FFT) are sufficient to extract detailed information about solid-body rotation states given data of high enough quality.

The flyby of Duende was unusual in that a full 1 year of advance notice was available. However such planetary encounters are not necessarily rare amongst NEOs. Since Duende flew by in 2013, eighteen other NEOs have passed by the Earth at less than 0.1 lunar distance or 6 Earth radii (though encounters with objects of Duende's size or bigger are less common). Such future planetary flybys, given enough advance notice, will provide opportunities to further test the phenomena explored here. Most notably, the 2029 encounter of the approximately 300-meter asteroid 99942 Apophis at a distance $<$6 Earth radii will be an excellent opportunity to test models predicting significant changes in that asteroid's rotation state \citep{scheeres05}. Observations of such flybys will further inform models for the currently unknown internal structure and cohesive properties of asteroids in the near-Earth population. As ongoing sky surveys continue to increase the annual yield of newly discovered near-Earth objects \citep[e.g.][]{christensen12,tonry11}, the rate of advance predictions of planetary flybys is increasing. This rate will grow even further as next generation surveys like the Large Synoptic Survey Telescope \citep[LSST;][]{jones09} come online. Experiments, like what we have presented here, to investigate processes of planetary geophysics in real-time will thus  become increasingly feasible.

\ack

We are extremely grateful to Marina Brozovi\'c and an anonymous reviewer for careful reading and thoughtful insights that significantly improved this manuscript. N.M. would like to acknowledge support from the National Science Foundation Astronomy and Astrophysics Postdoctoral fellowship program and the Carnegie Institution for Science, Department of Terrestrial Magnetism during the initial planning and executions stages of this campaign. Completion of the work was supported by NASA grant numbers NNX14AN82G and NNX17AH06G (PI N. Moskovitz) issued through the Near-Earth Object Observations program to the Mission Accessible Near-Earth Object Survey (MANOS). D.P is grateful to the AXA research fund for their postdoctoral fellowship. F.M was supported by the National Science Foundation under Grant Number 1743015. The observations presented here would not have been possible without the outstanding contributions from numerous telescope operators and support scientists, including David Summers, Daniel Harbeck and Ralf Kotulla at the WIYN 3.5m Observatory, Diane Harmer at the Kitt Peak 2.1m, Shai Kaspi at the Wise Observatory in Israel, and Chad Trujillo at Gemini. We are grateful to Lance Benner (JPL) and Michael Busch (NRAO) for helping guide us through the interpretation of their radar data. Beatrice Mueller kindly provided code for the WindowClean computation. Despite our efforts a number of planned observations were unsuccessful, we would like to acknowledge the efforts to observe made by Michelle Bannister, Ovidiu Vaduvescu, and Herve Bouy. We appreciate Bruce Gary making his lightcurve data publicly available. None of this work would have been possible without the invaluable resources provided by the Minor Planet Center and the JPL Horizons system. A. Stark is grateful for training received at the House of Black and White in the free city of Braavos. The paper is based in part on data obtained at Kitt Peak National Obseratory with the WIYN observatory and the Kitt Peak 2.1m. Kitt Peak National Observatory, National Optical Astronomy Observatory is operated by the Association of Universities for Research in Astronomy (AURA) under cooperative agreement with the National Science Foundation. The WIYN observatory is a join facility of the University of Wisconsin-Madison, Indiana University, Yale University, and the National Optical Astronomy Observatory. Data were also obtained with the VATT: the Alice P. Lennon Telescope and the Thomas J. Bannan Astrophysics Facility. This paper also includes observations obtained at Gemini-South Observatory, which is operated by the Association of Universities for Research in Astronomy, Inc., under a cooperative agreement with the NSF on behalf of the Gemini partnership: the National Science Foundation (United States), National Research Council (Canada), CONICYT (Chile), Ministerio de Ciencia, Tecnolog\'{i}a e Innovaci\'{o}n Productiva (Argentina), Minist\'{e}rio da Ci\^{e}ncia, Tecnologia e Inova\c{c}\~{a}o (Brazil), and Korea Astronomy and Space Science Institute (Republic of Korea). Gemini data were processed using the Gemini IRAF package. The data presented here were also obtained [in part] with ALFOSC, which is provided by the Instituto de Astrofisica de Andalucia (IAA) under a joint agreement with the University of Copenhagen and NOTSA. The ALFOSC observations were made with the Nordic Optical Telescope, operated by the Nordic Optical Telescope Scientific Association at the Observatorio del Roque de los Muchachos, La Palma, Spain, of the Instituto de Astrofisica de Canarias. This paper also uses observations made at the South African Astronomical Observatory (SAAO). Finally, this paper includes data gathered with the 100" Irenee du Pont telescope and the 6.5m Magellan Baade telescope, both located at Las Campanas Observatory, Chile and operated by the Carnegie Institution for Science.

\label{lastpage}





\end{document}